\documentclass[aps, prx,twocolumn,longbibliography,superscriptaddress,amsmath,amssymb,floatfix]{revtex4}
\usepackage[latin9]{inputenc}
\usepackage{amssymb}
\usepackage{graphicx}
\usepackage{amsmath}
\usepackage{color}
\usepackage{mathrsfs}
\usepackage{float}
\usepackage{indentfirst}
\usepackage{mathrsfs}
\usepackage{float}
\usepackage{indentfirst}
\usepackage{textcomp}
\usepackage{comment}
\usepackage{mathtools}
\usepackage{natbib,hyperref}
\usepackage{soul}

\begin{document}

\title{Stripe versus superconductivity in the doped Hubbard model on the honeycomb lattice}

\author{Mingpu Qin} \thanks{qinmingpu@sjtu.edu.cn}
\affiliation{Key Laboratory of Artificial Structures and Quantum Control,  School of Physics and Astronomy, Shanghai Jiao Tong University, Shanghai 200240, China}

\begin{abstract}
   We study the ground state of the doped Hubbard model on the honeycomb lattice in the small doping and strongly interacting region. The nature of the ground state by doping
 holes into the anti-ferromagnetic Mott insulating states on the honeycomb lattice remains a long-standing unsolved issue, even though tremendous efforts have been spent
 to investigate this challenging problem. 
 In this work, we employ two complementary, state-of-the-art, many-body computational methods -- constrained path (CP) auxiliary-field quantum
 Monte Carlo (AFQMC) with self-consistent constraint and density matrix renormalization group (DMRG) methods. Systematic and detailed cross-validations are performed between these two
 methods for narrow systems where DMRG can produce reliable results. AFQMC are then utilized to study wider systems to investigate the thermodynamic limit properties.
 The ground state is found to be a half-filled stripe state in the small doping and strongly interacting region. The pairing correlation shows $d$-wave symmetry locally, 
 but decays exponentially with the distance between two pairs. 
\end{abstract}

\maketitle

\section{Introduction} 
Understanding the physics of doped Mott insulator is one of the most important themes in condensed matter physics \cite{RevModPhys.78.17}.
It is now widely believed that the high-temperature superconductivity in cuprates is intimately related to the doping of a Mott insulator
on square lattice \cite{RevModPhys.78.17}.
Hubbard model (and its descendants) \cite{2021arXiv210312097A,2021arXiv210400064Q} is the minimum model to study Mott-related physics.  
The Hubbard model on honeycomb lattice is a prototype to study the correlated effect of electrons in two-dimensional materials with honeycomb structure like
graphene \cite{RevModPhys.81.109}.
It is an ideal model system to study the correlation-driven metal-insulator
transition \cite{RevModPhys.70.1039}. At half-filling, the Fermi surface shrinks to two
Dirac points with a linear dispersion on the honeycomb lattice. The ground properties at half-filling were accurately
determined by Quantum Monte Carlo (QMC) method without suffering from the infamous minus-sign problem because the honeycomb lattice
is bipartite \cite{Sorella_1992,Sorella_2012,PhysRevX.3.031010}.
A phase transition
occurs from the Dirac semi-metal phase
at weak interactions to the Mott insulator phase with long-range anti-ferromagnetic (AF) Neel order at strong interactions. 
The critical interaction strength is determined to be $U_c \approx 3.8$ \cite{PhysRevX.3.031010,PhysRevX.6.011029} and
the phase transition is found to be in the Gross-Neveu-Yukawa \cite{PhysRevX.3.031010,PhysRevX.6.011029} universality class.

A definite answer to how the AF ground state on the honeycomb lattice in the strongly interacting region evolves with doping in the system is still lacking.
Numerically, the infamous minus sign problem
emerges when the system is doped away from half-filling which hampers the investigation of large system sizes at low temperature with QMC \cite{PhysRevB.41.9301,PhysRevLett.94.170201}.
There exists approaches which don't suffer from the sign problem but have other difficulties \cite{doi:10.1146/annurev-conmatphys-020911-125018,PhysRevX.5.041041}. 
In general, the competition between kinetic and
potential energies can lead to exotic states when holes are introduced into the Mott insulator \cite{ANDERSON1196}.

Historically, the doped Hubbard model on the honeycomb lattice was extensively studied and different candidates for the ground state were proposed.
The one-quarter doping case has attracted tremendous attention because the density of states shows a Van Hove singularity and 
the Fermi surface is nested, which usually triggers instabilities towards different types of orders.
At weak interaction, which is relevant to graphene \cite{RevModPhys.81.109}, $d+id$ superconductivity was predicted in the Hubbard model on the honeycomb lattice near one-quarter doping
by different methods \cite{PhysRevB.81.085431,PhysRevLett.100.146404,Nandkishore_2012,PhysRevB.86.020507,PhysRevB.81.224505,PhysRevB.85.035414,PhysRevX.4.031040}.
Spontaneous quantum Hall effect was also found at one-quarter doping \cite{Li_2012,PhysRevB.85.035414,PhysRevX.4.031040}. In this work, we focus on the strongly interacting region where correlation effect
plays an essential role. Chiral $d+id$ superconductivity was also predicted in the strongly interacting region. For example, in \cite{PhysRevB.90.054521}
and \cite{PhysRevB.88.155112},
chiral $d + id$ superconductivity was found by QMC and by a tensor network states related method (in the large $U$ limit, i.e., the t-J model) respectively.
In a recent work, $p+ip$ superconductivity was obtained with Grassmann tensor product state approach \cite{PhysRevB.101.205147} in the t-J model. 

Experimentally, long-range AF Neel order, which is the ground state
in the strongly interacting region without doping, was observed in ${\text{Na}}_{2}{\text{IrO}}_{3}$ \cite{PhysRevB.82.064412}
and $\text{InCu}_{\frac{2}{3}}\text{V}_{\frac{1}{3}}\text{O}_3$ \cite{KATAEV2005310}. Superconductivity was also discovered in the pnictide SrPtAs with a honeycomb structure \cite{doi:10.1143/JPSJ.80.055002}, in which time-reversal symmetry was found to be broken \cite{PhysRevB.87.180503}. 
A pressure-driven superconductivity in $\text{FeP}\text{Se}_3$ with an iron-based honeycomb lattice structure was reported recently \cite{Wang18}. 

In this work we study the ground state properties of doped Hubbard model on the honeycomb lattice in the strongly interacting and lightly doped region. We employ two complementary, state-of-the-art, many-body computational methods -- constrained path (CP) auxiliary-field quantum
Monte Carlo (AFQMC) with self-consistent constraint and density matrix renormalization group (DMRG) methods.
We perform detailed cross-validation for width-4 cylinders where DMRG can give very accurate results and then
study wider systems with AFQMC to obtain the thermodynamic properties. We calculate the distribution of the doped holes on the lattice and the evolution
of the AF order with doping. To detect the possible superconducting order, we calculate the pair-pair correlation functions
and analyze both the long-range behavior and the local structure.

We find a half-filled stripe order in the vicinity of half-filling with strong interaction, i.e., $1/16$ and $1/12$ dopings with $U = 8$.
This stripe state was previous obtained in width-4 cylinder at $1/16$ doping \cite{PhysRevB.103.155110}. But in this work we
employ two complementary methods and study wider systems. We find the pair-pair correlation functions decay exponentially
with the distance between two pairs which indicates the absence of long-range pairing order in the system. We also analyze the pairing symmetry and find the pair-pair correlation
displays a $d$-wave symmetry locally. The stripe phase is found to terminate around $1/8$ doping at $U = 8$ without the emergence of long-range pairing order. 
These results indicate a complicated relationship between stripe order and superconductivity in the doped Hubbard model on the honeycomb lattice.

\begin{figure}[t]
	\includegraphics[width=40mm]{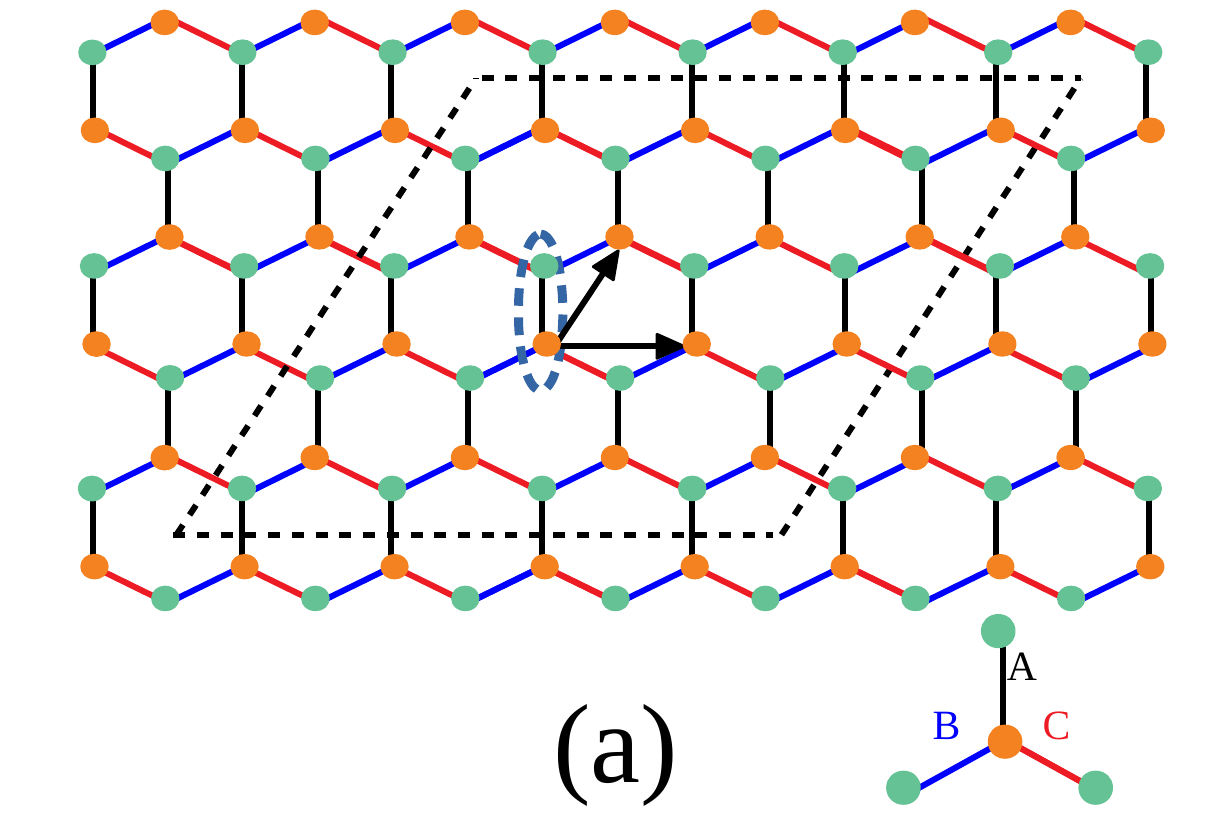}
	\includegraphics[width=44mm]{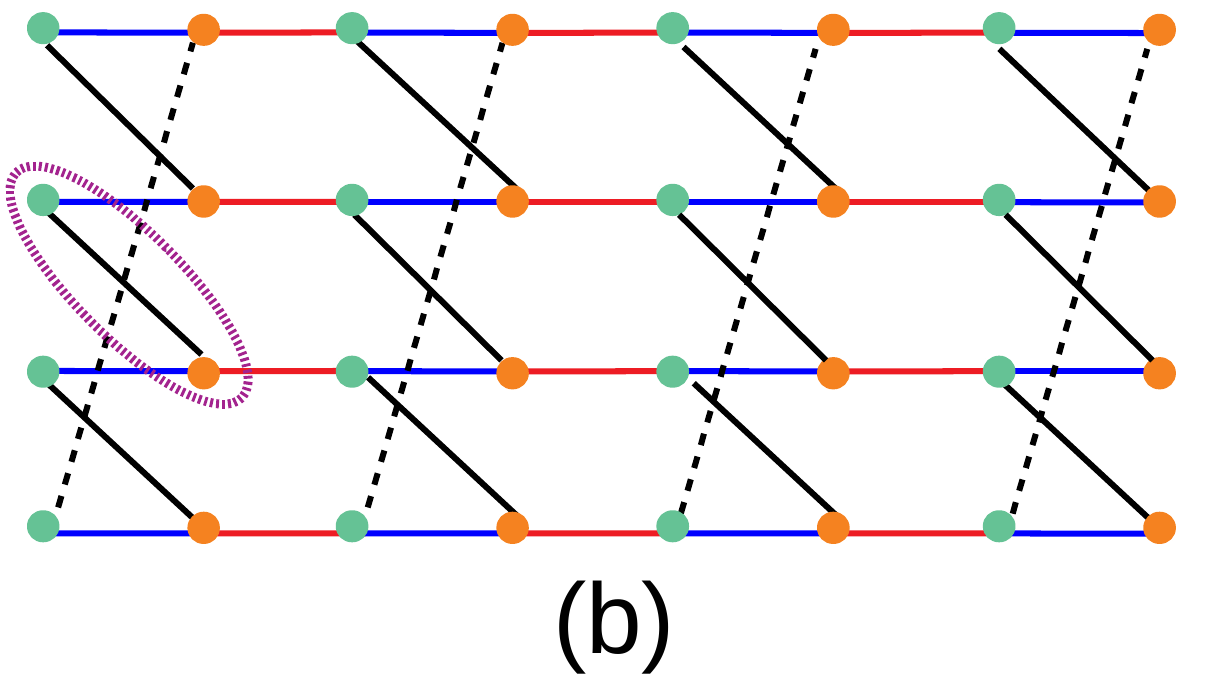}
	\caption{(a) An illustration of the honeycomb lattice. Each unit cell consists of two sites as denoted by
		the green and orange dots in the dashed oval. Bonds along different directions are distinguished with
		colors and are labeled as A, B and C. The two arrows are the primitive vectors of the Bravais lattice. (b) The $4 \times 4$ super cell in (a)
		is rearranged into a $8 \times 4$ square lattice with only horizontal bonds and next nearest bond in the upper left
		(or lower right) directions.
		In the DMRG calculation, periodic (open) boundary conditions are imposed for the vertical (horizontal) directions.
		The dash lines represent the interactions due to the periodic boundary conditions. The A bond in the dashed oval is the
		reference bond when calculating the pair-pair correlation function.
	} 
	\label{lattice}
\end{figure}

\begin{figure}[t]
	\includegraphics[width=80mm]{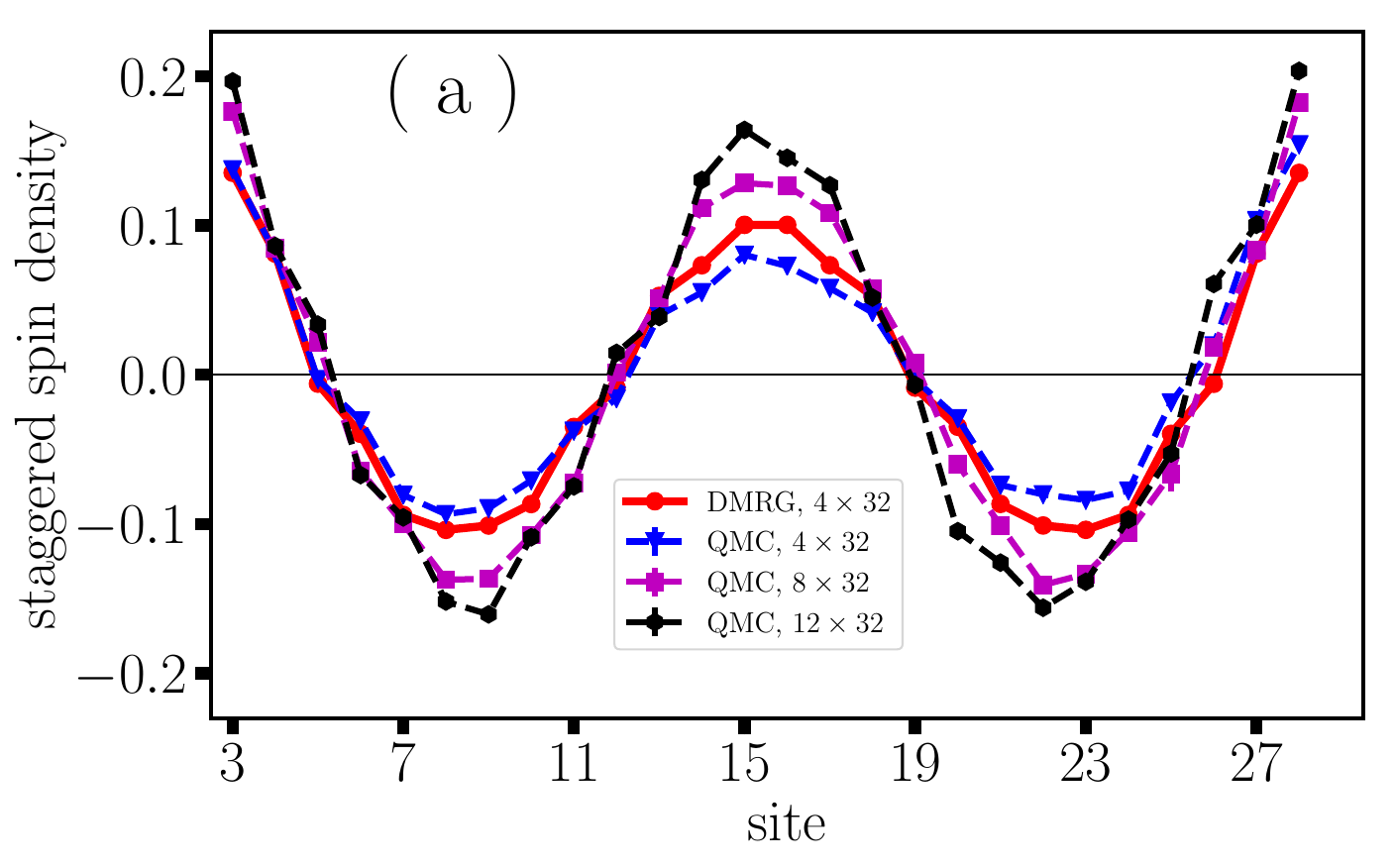}
	\includegraphics[width=80mm]{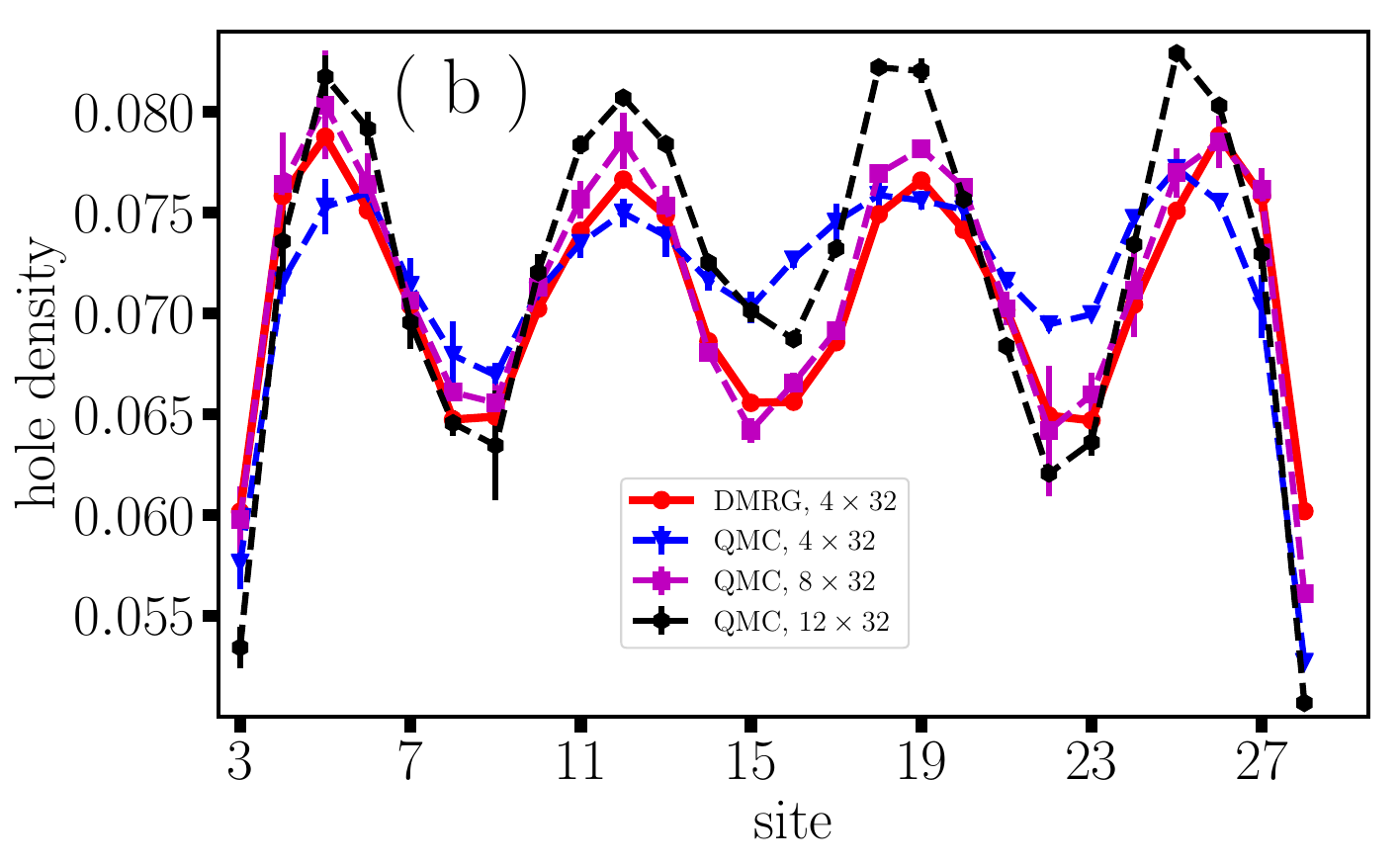}
	\caption{Staggered spin (a) and hole (b) density for system with length $32$ in the rearranged square lattice with $U = 8$
		at $h = 1/16$ doping.
		The DMRG results (red) for width $4$ cylinder are obtained from an extrapolation with truncation error. They are used as
		benchmark values for AFQMC results. The thin horizontal line in (a) represents zero.  A good agreement between AFQMC (blue) and DMRG results (red) for the $4 \times 32$ system
		can be seen. Results for wider cylinders from AFQMC (which are beyond the capacity of DMRG) are also plotted. These results
		show the ground state is a half-filled stripe state.} 
	\label{spin_charge}
\end{figure}

\section{Model and Computational methods}

\subsection{Model}
The Hamiltonian of the Hubbard model is
\begin{equation}
	H=-t\sum_{\langle i,j\rangle,\sigma}c_{i\sigma}^\dagger c_{j\sigma}+U\sum_{i}n_{i\uparrow}n_{i\downarrow}
\end{equation}
where $t$ is the hopping constant and it is set to the energy unit. We study the Hubbard model with strong repulsive interactions with $U = 8$ in this work.
We only consider spin-balanced case with equal number of electrons with up and down spin.
The local spin and hole density at site $i$ are $S_{i} = (\langle n_{i,\uparrow} \rangle - \langle n_{i, \downarrow} \rangle) / 2$ and
$h_i = (1 - \langle n_{i,\uparrow} \rangle - \langle n_{i, \downarrow} \rangle)$ respectively. 

An illustration of the honeycomb lattice is shown in Fig.~\ref{lattice}. In this work, we rearrange the honeycomb lattice into a square one
in order to index the sites more conveniently. In Fig.~\ref{lattice}, the $4 \times 4$ supercell is rearranged into a $8 \times 4$ square lattice 
with only horizontal bonds and next nearest bonds in the upper left (or lower right) directions. Throughout this work, we index
the sites using coordination pair $(x, y)$ following the convention of the square lattice. It is worth noting that there is a scaling factor
between the measurement of distance on square lattice and on the original honeycomb lattice. Nevertheless, this factor doesn't affect the conclusions in this work, e.g,
when discussing the decay of pair-pair correlations with distance.

We study systems with cylinder geometry, i.e., with periodic (open) boundary conditions in vertical (horizontal) directions, to favor the DMRG
calculation. Anti-ferromagnetic magnetic pinning fields are applied at the edges of cylinder so we can measure the local spin and hole densities instead of the
more demanding correlation functions to probe the stripe order \cite{PhysRevLett.99.127004}.
We plot the staggered spin density, $(-1)^{i_x}S_i$ from which the stripe structure is easier to identify.

We probe the possible pairing order by measuring the pair-pair correlation $\langle \hat{\Delta}_{i'j'}^{\dagger} \hat{\Delta}_{ij} \rangle$ with
singlet pairing operator defined as $\hat{\Delta}_{ij}\equiv(\hat{c}_{i\uparrow}\hat{c}_{j\downarrow}-\hat{c}_{i\downarrow}\hat{c}_{j\uparrow})/{\sqrt{2}}$.
We find that the correlation in the triplet channel is weaker than the singlet correlation, so we only show the results for singlet correlation in this work.

\subsection{Density Matrix Renormalization Group}


DMRG \cite{PhysRevLett.69.2863} was developed by considering the effect of environment
in the renormalization process. It is extremely accurate for one-dimensional (1D) quantum systems and is now arguably the workhorse for 1D problems.
The success of DMRG lies in the underlying MPS wavefunction which captures the entanglement structure of 1D systems. Despite the difficulty
in the application of it to two-dimensional systems, DMRG has played an essential role in the study of narrow cylinders for which relatively accurate
results can be obtained by pushing the kept states in DMRG to tens of thousands \cite{PhysRevX.10.031016}. In this work, we use DMRG to study width-4 cylinder.
The kept state in our calculation is as large as $10000$. Extrapolation with truncation errors are performed to remove the
finite kept state effect.  


\subsection{Auxiliary-field quantum Monte Carlo}
In AFQMC calculation the interacting (two-body) terms are represented as an ensemble of one-body term fluctuating in the auxiliary bosonic field through
the Hubbard-Stratonovich decomposition.
Then classical Monte Carlo techniques are employed to evaluation physic quantities which are basically ultrahigh-dimensional integral (summation). With only a few exceptions, QMC
suffers from the infamous negative sign problem \cite{PhysRevB.41.9301} which hampers the study of systems with large size or at low temperature.  One strategy to overcome
the negative sign problem is to take advantage of the bias-variance trade-off. We can get rid of the sign problem by modify the sample process in AFQMC. But the price to
pay is the introduction of bias in the results. Constrained-path (CP) AFQMC was developed under this spirit \cite{Zhang97}, in which a trial wave-function is introduced to
control the sign problem by discarding samples whose overlaps with trial wave-function are negative. 
In a recent advance, an iteration process is augmented with CP-AFQMC to optimize the trial wave-function and reduce the bias,
making the
method self-consistent \cite{PhysRevB.94.235119}. CP-AFQMC augmented with this new gradient played an important role in the determination of the stripe state in the
doped Hubbard model on
square lattice \cite{Zheng1155}.  
As we will discuss below, for width-4 cylinder where DMRG are reliable, results from CP-AFQMC with self-consistently optimized trial wave-function agree well with DMRG
values. AFQMC are then employed to study wider cylinders which are beyond the capacity of DMRG, to obtain the thermodynamic properties.

\section{Cross validation and Half-filled stripe order}

\begin{figure*}[t]
	\includegraphics[width=59mm]{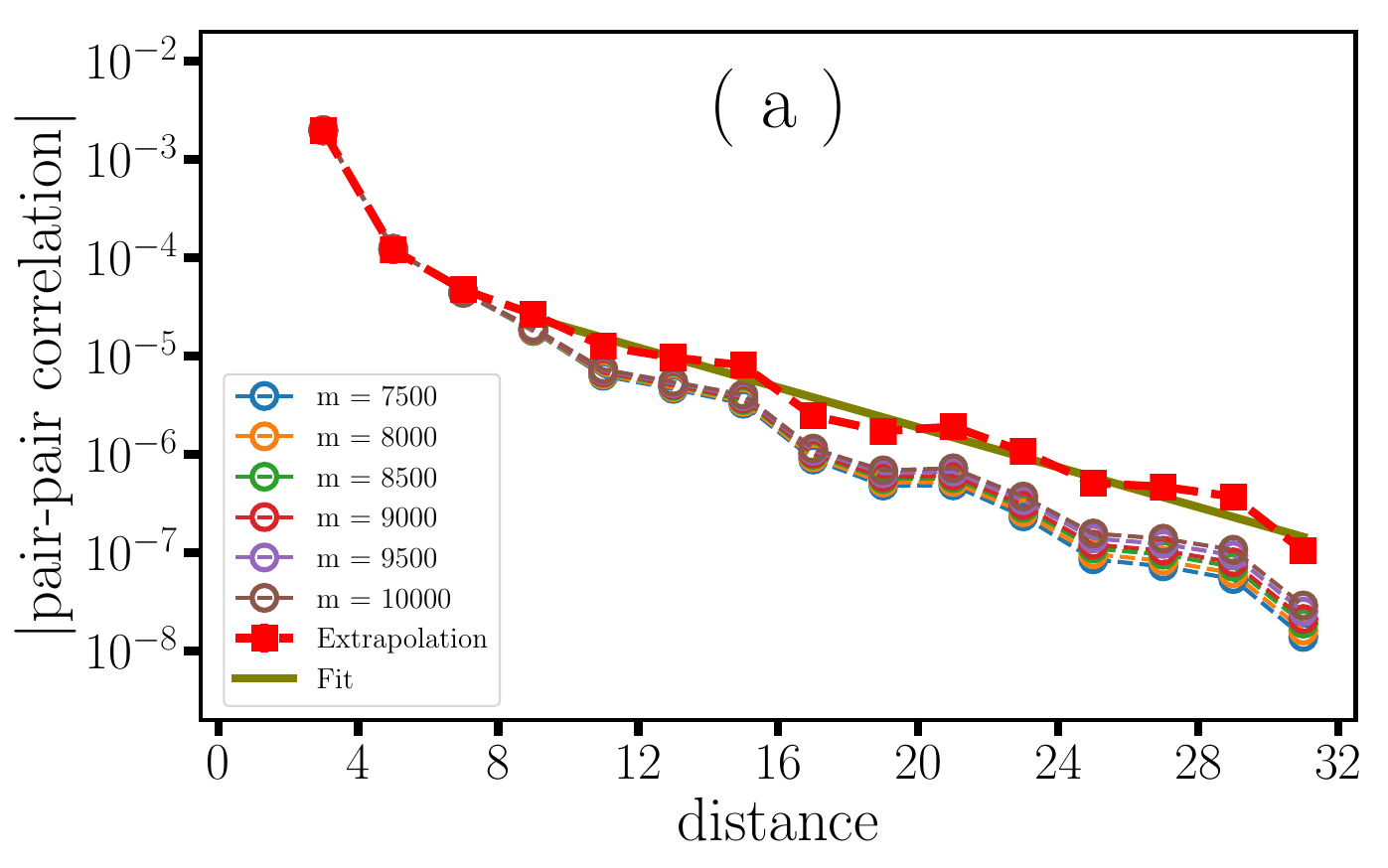}
	\includegraphics[width=59mm]{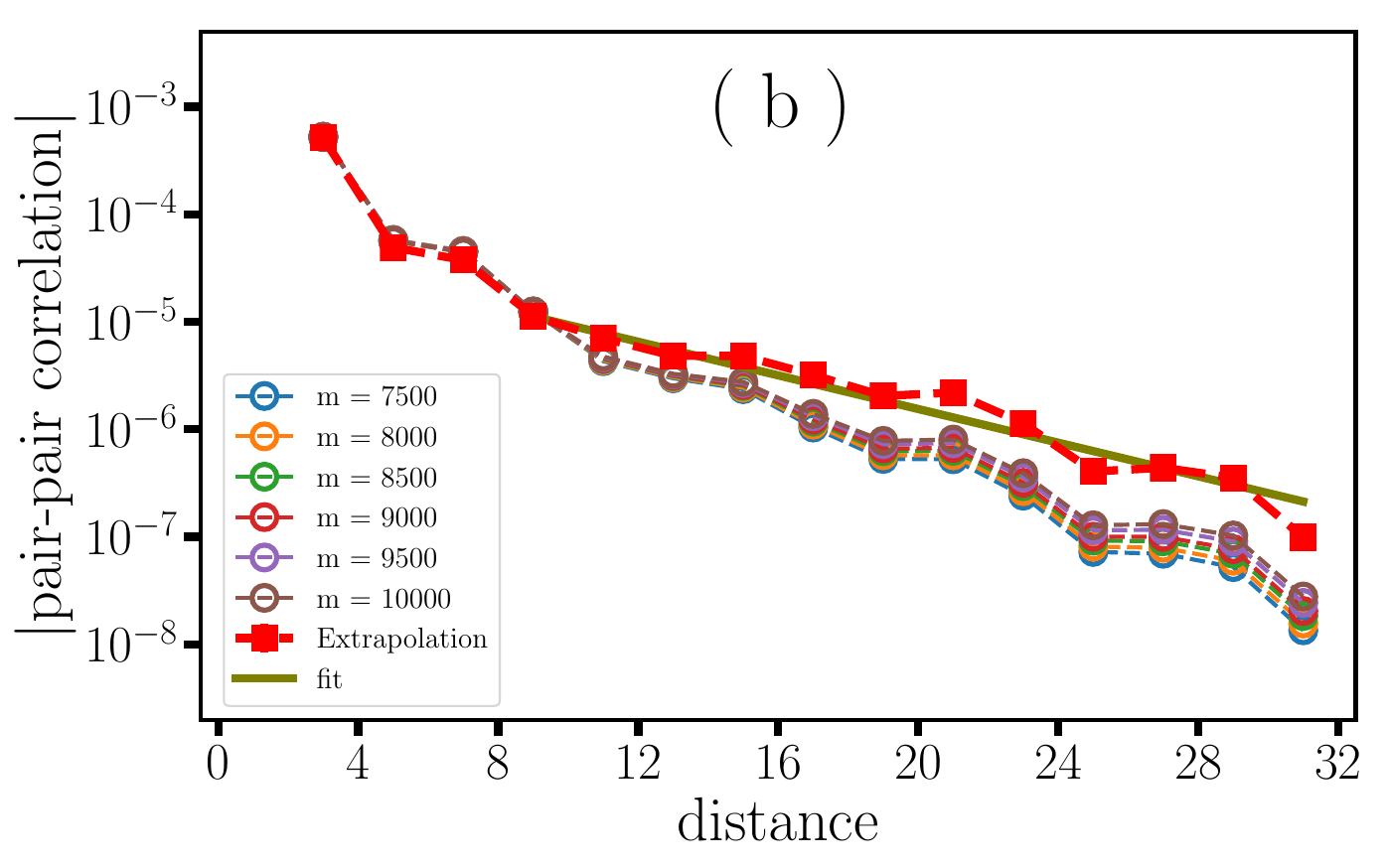}
	\includegraphics[width=59mm]{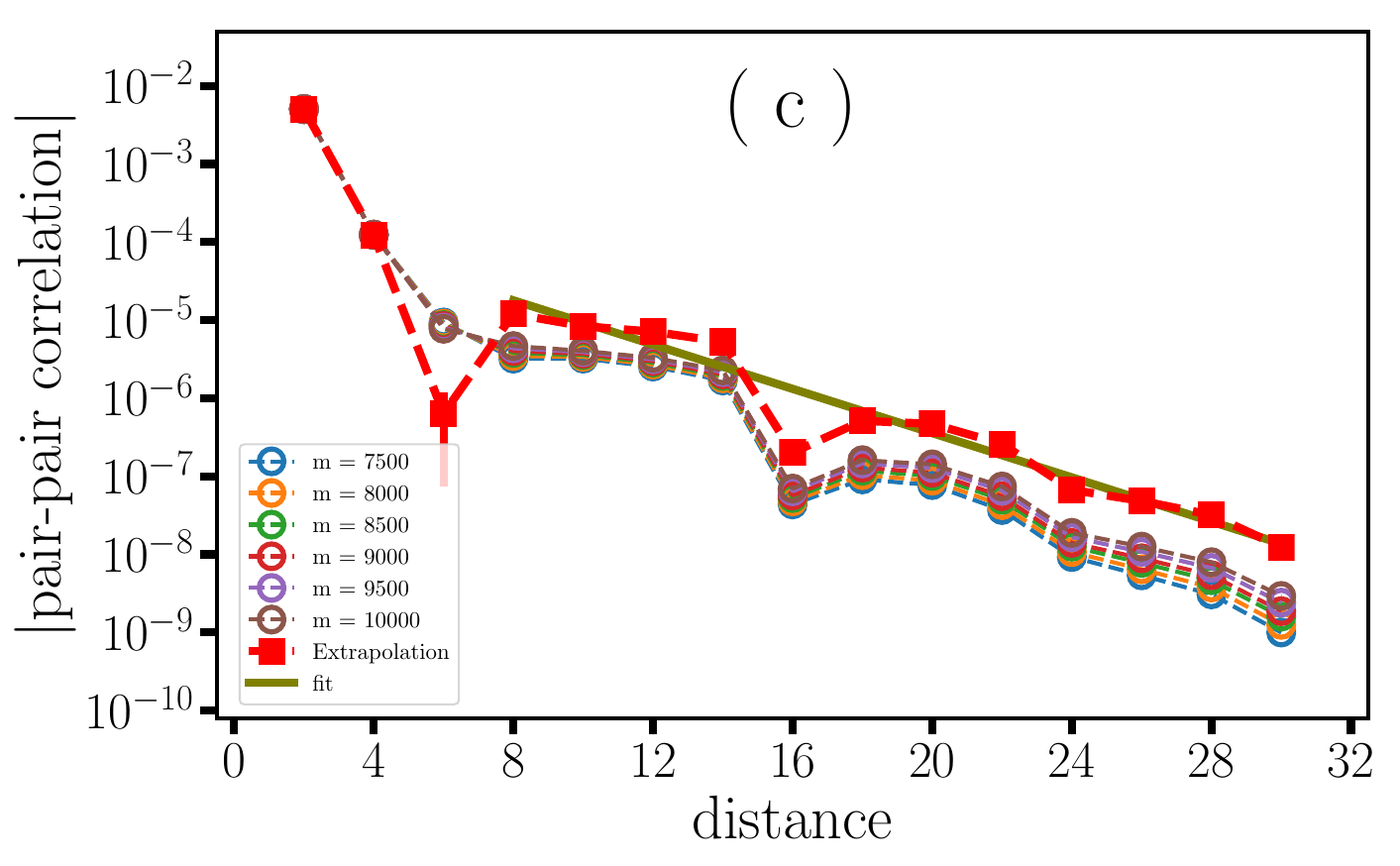}
	\caption{DMRG results for the absolute value of pair-pair correlations for $4 \times 32$ system with $U = 8$ and $1/16$ doping. The reference bond is placed at the
		edge between site $(1, 3)$ and $(2, 2)$. Panels (a), (b), and (c) show the correlation versus distance between the reference bond (The A bond in the dashed oval
		in Fig.~\ref{lattice}) and the
		black (A), blue (B), and red (C) bonds (see Fig.~\ref{lattice}) respectively. Both results with finite kept states and from an extrapolation (red) with truncation errors are shown.
		An exponential decay can be seen from the fit (brown).} 
	\label{pair_pair}
\end{figure*}

We first study the $4 \times 32$ system at $1/16$ doping with $U = 8$. We apply AF magnetic pinning fields with strength $|h_m| = 0.5$ at the
open edges of the cylinder to enable the probe the spin order by measuring the local spin density.
For width-4 cylinder, DMRG can provide accurate results which enable a benchmark of AFQMC
calculations. The spin and hole density from DMRG are depicted as the red lines in Fig.~\ref{spin_charge}.
We only plot the result for one row because all the other rows have the same values due to the PBC in the vertical direction. DMRG calculations are performed
with kept state $m$ as large as $10000$. We only show the extrapolated to zero truncation error
results and the details of the extrapolation process can be found in the Appendix. We can clearly see a stripe \cite{PhysRevB.40.7391} structure in Fig.~\ref{spin_charge} in which
the holes are concentrated
at the place where spin density display a node ($\pi$ phase flip). The stripe state is half-filled since there are $4$ stripes and $8$ holes totally at $1/16$ doping, while the width of the system is $4$.

Converged AFQMC results with self-consistently optimized trial wave-function for the same $4 \times 32$ system are also shown as the blue dotted line in Fig.~\ref{spin_charge}.
The details of the optimization process can be found in the Appendix.
We can see the spin density from AFQMC agrees very well with DMRG result. There exists tiny discrepancy for hole density but the stripe structure is the same,
i.e., the holes are concentrated at the node place of staggered spin density.
These results from AFQMC agree well with the conclusion in previous studies \cite{PhysRevB.94.235119} where it was found that AFQMC with self-consistent constraint provides
very accurate
results for spin density. The ground state energies from DMRG and AFQMC are $-73.081(4)$ (extrapolated to zero truncation error) and $-73.22(1)$ respectively with a
relative error of AFQMC less than $-0.2\%$ which also matches the previous conclusion \cite{PhysRevB.94.235119}.  

We then employ AFQMC to calculate wider cylinders which are beyond the capacity of DMRG. In Fig.~\ref{spin_charge}, we show the spin and hole density
of cylinders with size $8 \times 32$ and $12 \times 32$ and $U = 8$ at the same $1/16$ doping. For these systems, we use the unrestricted Hartree Fock trial wave-function
with an effective $U = 3$ same as
the converged effective $U$ values in the self-consistent process for $4 \times 32$ system \cite{PhysRevB.94.235119} (see the Appendix). We can find an increase
of the amplitude of the modulation for
both the spin and hole density with the
increase of the width of the cylinder from $4$ to $8$. The spin and hole density are nearly converged at width $8$ comparing to the width $12$ results. These results establish the half-filled stripe order
in the ground state of $1/16$ doped Hubbard model on the honeycomb lattice in the 2D thermodynamic limit.

We also study $1/12$ doping, $U = 8$ case with DMRG for width-$4$ cylinder and observe the same half-filled stripe ground state but with a weaker order. We find that the stripe phase terminates
at $1/8$ doping with $U = 8$. The details of these results
are presented in the Appendix.

\begin{figure*}[t]
	\includegraphics[width=180mm]{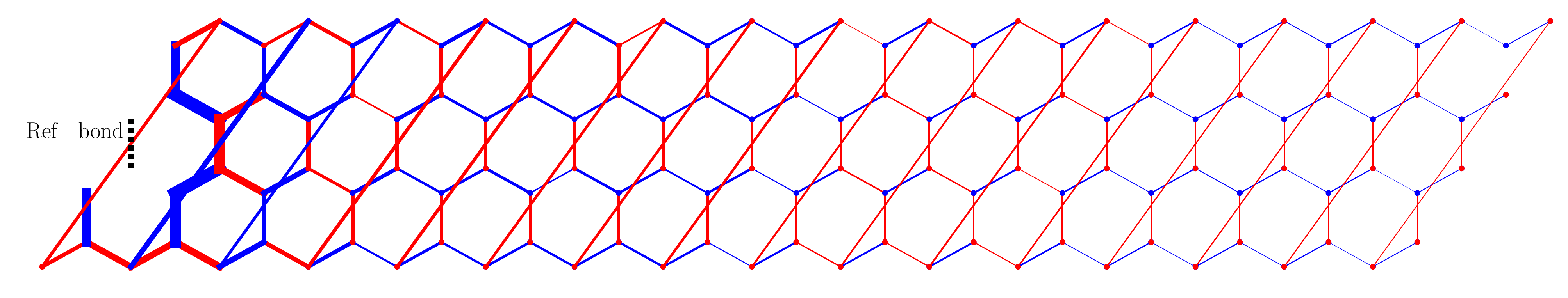}
	\caption{The pair-pair correlation pattern on the whole lattice. The reference bond is denoted by the dashed black line. Red (blue) color
		means positive (negative) correlation values. The thickness of each bond is proportional to  $\langle \hat{\Delta}_{i'j'}^{\dagger} \hat{\Delta}_{ij} \rangle ^\frac{1}{4}$ 
		to make the line
		visible. We don't show the results for bonds sharing site with the reference bonds because there is contribution from local density for the correlations of these bonds.
		We can see that the A (B) bonds (definition in Fig.~\ref{lattice}) have positive (negative) correlation at large distance, while the sign of C bonds oscillate with
		the distance to the reference bond. The relative strengths of the bonds connected by the same sites are plotted in Fig~\ref{pair_ratio}.} 
	\label{pair_color}
\end{figure*} 

\section{Superconducting pairing}
The stripe order could intertwine with superconductivity and the coexistence of them can results in the so called pair density wave states \cite{Fradkin15}.
To probe the possible coexisting superconductivity we calculate the pair-pair correlation function with DMRG.


\subsection{Pair-pair correlation}
In Fig.~\ref{pair_pair}, we plot the pair-pair correlation function for the $4 \times 32$ cylinder at $1/16$ doping
with $U = 8$. The reference bond is the A bond placed at the
edge of the cylinder between sites $(1, 3)$ and $(2, 2)$ (see Fig.~\ref{lattice}). In the three panels (a), (b), and (c), we plot the absolute value of correlation for
black (A), blue (B), and red (C) bonds in Fig.~\ref{lattice} respectively. DMRG results with kept states from $m = 7500$ to $10000$ are shown. We also
perform an extrapolation with truncation error and the extrapolated results are denoted by the red lines (details in the Appendix). 
The brown lines are exponential fits of the extrapolated values, from which we know the pair-pair
correlation decays exponentially with the distance between two pairs.
We can also see a tiny oscillation in the exponential decay of the pair-pair correlations which was caused by the stripe order.
From these results we conclude that no superconductivity coexists with the stripe order in the ground state of the system.

We also study the $1/12$ and $1/8$ dopings, $U = 8$ cases with DMRG and find the pair-pair correlation also decays exponentially with distance.
The details can be found in the Appendix.

\subsection{Local pairing symmetry}
In Fig.~\ref{pair_color}, we plot the sign structure of the pair-pair correlation function for each bond.
In Fig.~\ref{pair_color}, red (blue) color bonds have positive (negative) correlation with the reference bond
denoted as the black dashed line. The thickness of each red and blue bond represents the absolute value of the correlation. To
make the line visible to eyes, we set the thickness of each bond proportional to $\langle \hat{\Delta}_{i'j'}^{\dagger} \hat{\Delta}_{ij} \rangle ^\frac{1}{4}$ on each bond.
We can see that
at long distance, all the A bonds (see Fig.~\ref{lattice} for definition, same for following discussions) have positive correlations, while the correlations for B bonds are negative.
The C bonds have the weakest correlation and the sign oscillates with the distance.   

\begin{figure}[b]
	\includegraphics[width=80mm]{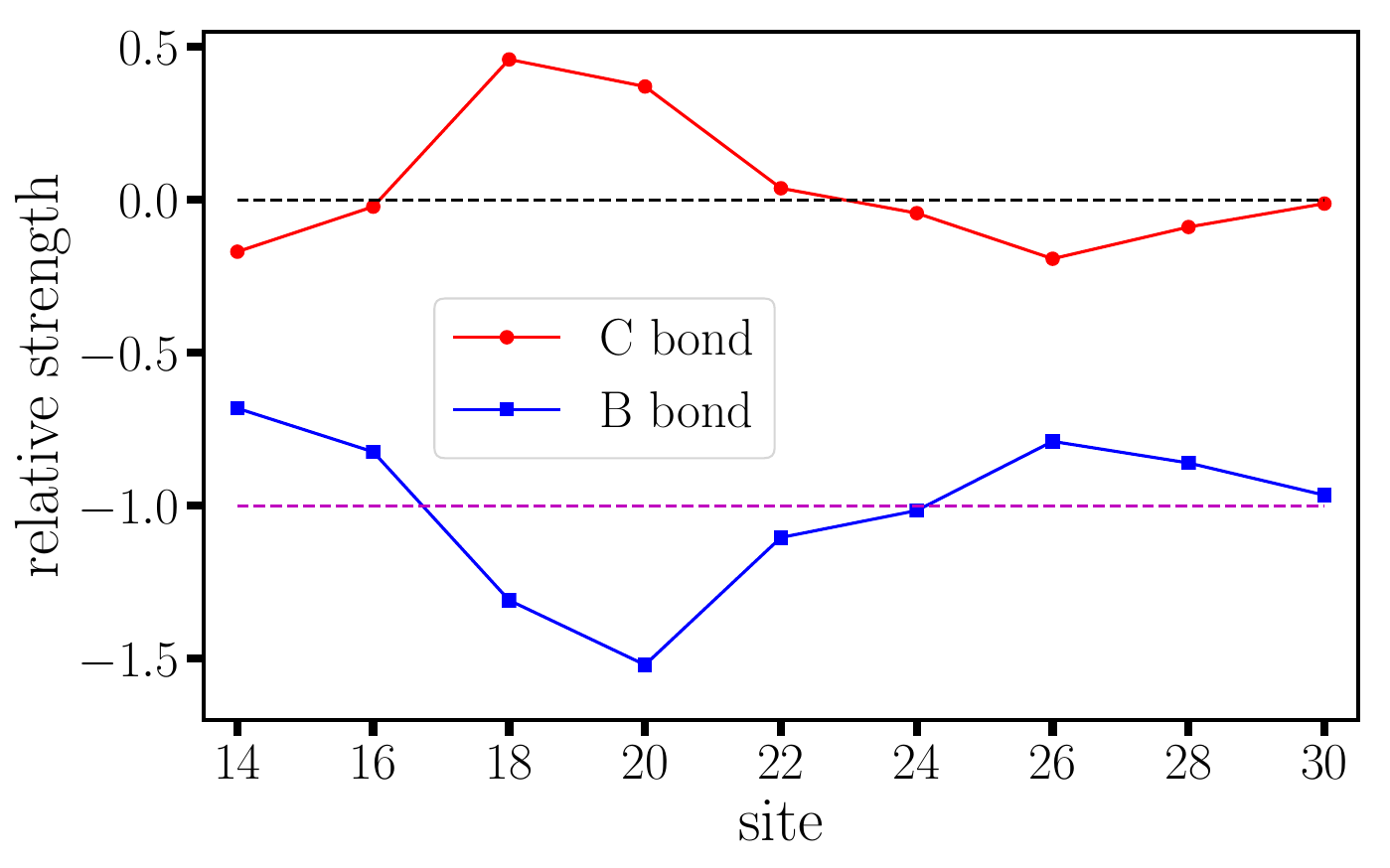}
	\caption{The relative strengths of the pair-pair correlations for bonds connected by the same sites. We set the correlation of A bond at each site as reference.
		The dashed horizontal lines represent $0$ and $-1$.
		We can see that at large distance, the relative strength of pair-pair correlation for three bonds connected by the same site have a $(1,-1,0)$ structure approximately,
		which is a $d$-wave representation of the $D_{6h}$ symmetry group of the
		honeycomb lattice \cite{PhysRevB.75.134512}.} 
	\label{pair_ratio}
\end{figure}

To show the relative strength of A, B, and C bonds connected at the same sites, we plot the ratio of them in Fig.~\ref{pair_ratio}.
We divide the correlation on B and C bonds with the value on A bonds which are connected by the same sites. From Fig.~\ref{pair_ratio} we
can find that the strength of correlations on B bonds is nearly equal to the values on A bonds locally but with opposite sign. While the
correlations on C bonds are very tiny comparing to the values on A bonds. From these, we know the
pairing order have an approximate $(1, -1, 0)$ structure locally in the region far from the reference bond, which is exactly one of the degenerate
$d$-wave representations $d_{xy}$ of $D_{6h}$ symmetry of
the honeycomb lattice \cite{PhysRevB.75.134512}. Because the cylinder geometry we adopted doesn't preserve the $D_{6h}$ symmetry rigorously, the numerical results for correlation
select the $d_{xy}$ representation $(1, -1, 0)$ with small discrepancy to the exact ratios.     

We want to emphasize that although the pair-pair correlation shows $d$-wave symmetry locally, they decay exponentially with distance
as shown in Fig.~\ref{pair_pair} which indicates the
absence of long-range pairing order in the system.

\section{Comparison with the square lattice}
Honeycomb lattice is very similar to the square lattice
in the sense that they are both bipartite and both develope AF long-range order at strongly interacting region at half-filling.
But at half-filling, the Fermi surface of the honeycomb lattice shrinks to two Dirac points which cause the AF
Neel order to develop only when the interacting strength is larger than a finite critical $U_c \approx 3.8$ \cite{PhysRevX.3.031010}, while on square 
lattice $U_c = 0$ \cite{PhysRevB.94.085140}. 
On the square lattice, it is now established that the ground state of
the Hubbard model with only nearest hopping is a filled stripe without superconductivity in the vicinity of region with $1/8$ doping and
$U = 8$ \cite{Zheng1155,PhysRevX.10.031016}. The ground state of
doped Hubbard model on the honeycomb lattice is similar to that on the square lattice: a half-filled stripe state without superconductivity.
We notice that the stripe observed in this work on the honeycomb lattice is diagonal if we rearrange the honeycomb lattice into a brick wall
square lattice. Previous study shows diagonal filled stripe is very close to the true ground (filled vertical stripe) in energy for the t-J model on square lattice
\cite{PhysRevLett.113.046402}.
We also notice that the stripe order on the honeycomb lattice is weaker than that on square lattice, due to the larger quantum fluctuation on the honeycomb lattice
because of the smaller coordination number. And for the same reason, the critical doping where the stripe order disappears on the honeycomb lattice is larger than
that on the square lattice \cite{foot2} with $U = 8$. 

Intuitively, in stripe state, the AF background is preserved and the hole can also move ``freely" along the stripe which means a gain in kinetic energy.    
The appearance of stripe state on both the square and honeycomb lattices leads us to ask whether stripe phase is
a universal consequence when doping an AF Mott insulator. 

\section{Summary and perspectives}
In this work we investigate the ground state properties of the doped Hubbard model on the honeycomb lattice with two state of arts numerical methods.
We perform detailed cross-validation and find agreement between the two methods. We discover the half-filled stripe
order in the lightly doped and strongly interacting region which terminates around $1/8$ doping with $U = 8$. We find no long-range pairing order in the ground state. But the pair-pair
correlation displays a $d$-wave symmetry locally. 
The half-filled stripe order could be measured experimentally \cite{nature_375_15_1995} in real materials with honeycomb structure,
artificially synthesized honeycomb systems \cite{2013NatNa...8..625P}, or ultra-cold atom platform \cite{Parsons1253} with advance in
cooling technology. 
 Since these experiment are usually carried out at finite
temperature, it is also necessary to study how and when the stripe order melts with thermal fluctuation \cite{2020arXiv200910736W}.
The establishment of stripe provides a new beginning for the theoretical pursuit of superconductivity on the honeycomb lattice.
It will be interesting to study whether superconductivity could emerge by frustrating the stripe order with longer-range hoppings or interacting terms.

\begin{acknowledgments}
This work is supported by a start-up fund from School of Physics and Astronomy in Shanghai Jiao Tong University. 
We thank useful discussions with T. Xiang and C.-M. Chung, and thank C.-M. Chung, Schollw\"{o}ck, S. R. White and S. Zhang for earlier collaborations on related topic. 
The DMRG calculations
in this work are performed with the iTensor package \cite{itensor}.
\end{acknowledgments}

\bibliography{Honeycomb.bib}

\appendix

\section{$U = 8$, $1/16$ doping results}
\subsection{Extrapolation of DMRG results}

\begin{figure}[b]
	\includegraphics[width=80mm]{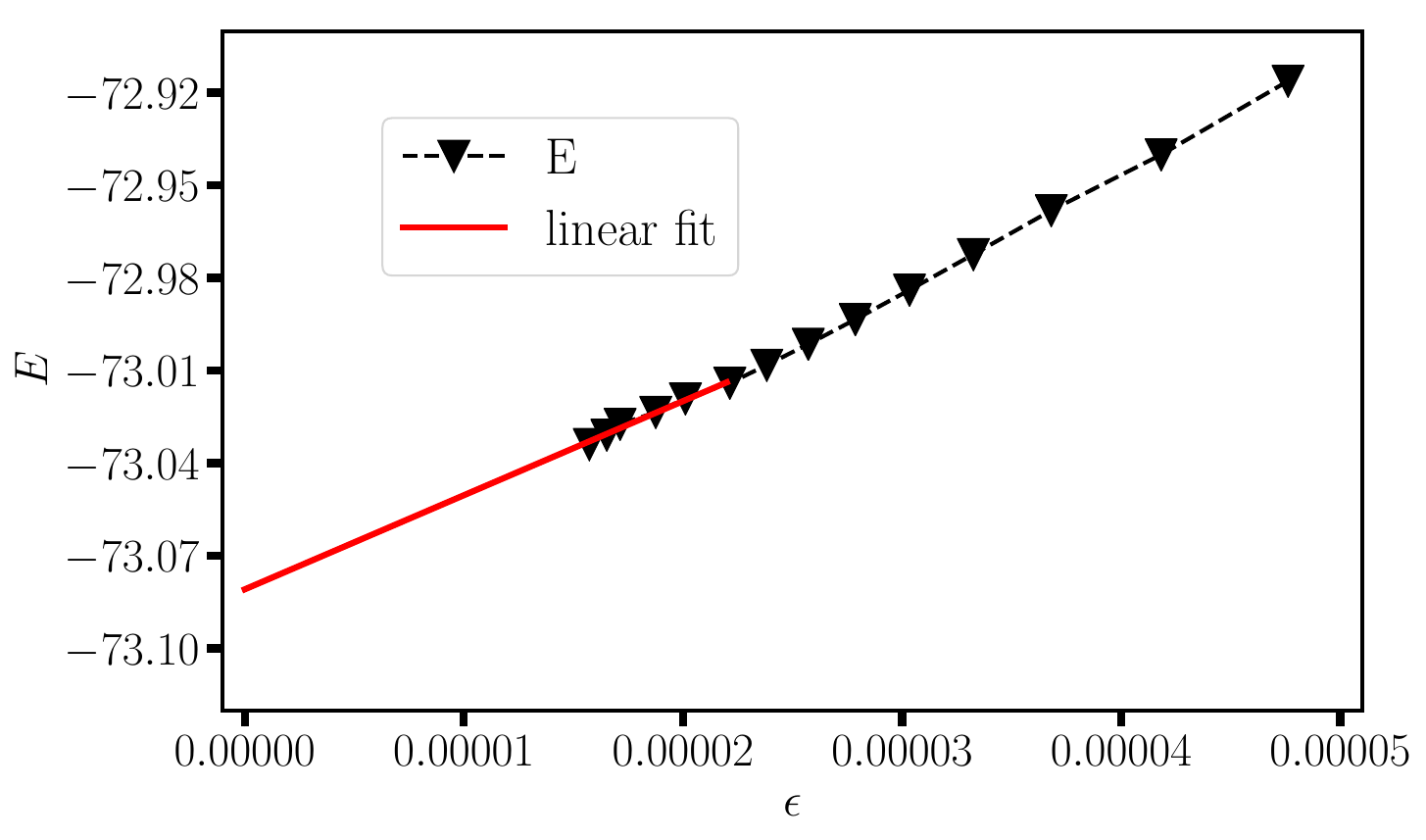}
	\caption{Scaling of energy with truncation error in DMRG calculation. The system is with size $4\times 32$, $U = 8$, and $1/16$ doping.
		The ground state energy is estimated to be $-73.081(4)$ with a linear extrapolation using the $6$ data with smallest truncation errors.} 
	\label{E_scale_4_32}
\end{figure}

In Fig.~\ref{E_scale_4_32}, we show the scaling of energy with the truncation error for the $4\times 32$ system with $U = 8$, and $1/16$ doping in DMRG calculation.
AF magnetic pining fields with strength $|h_m| = 0.5$ are applied at the open edges. 
With an extrapolation of the $6$ energies with the smallest truncation errors, the ground state energy is estimated to be $-73.081(4)$.  

\begin{figure}[t]
	\includegraphics[width=80mm]{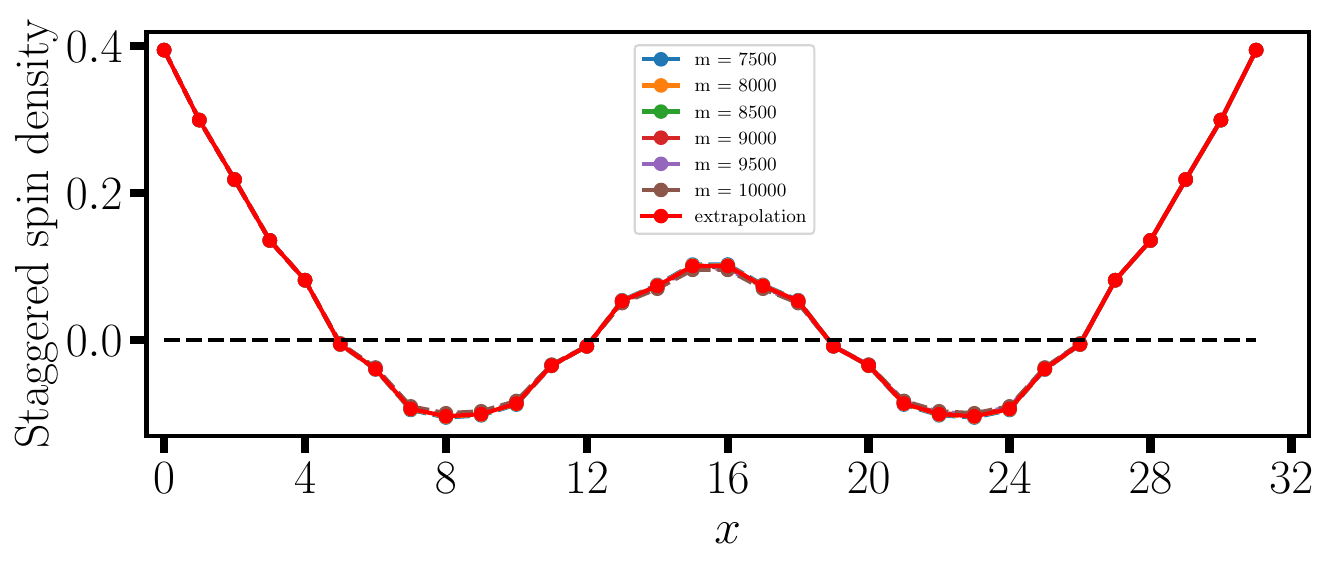}
	\includegraphics[width=80mm]{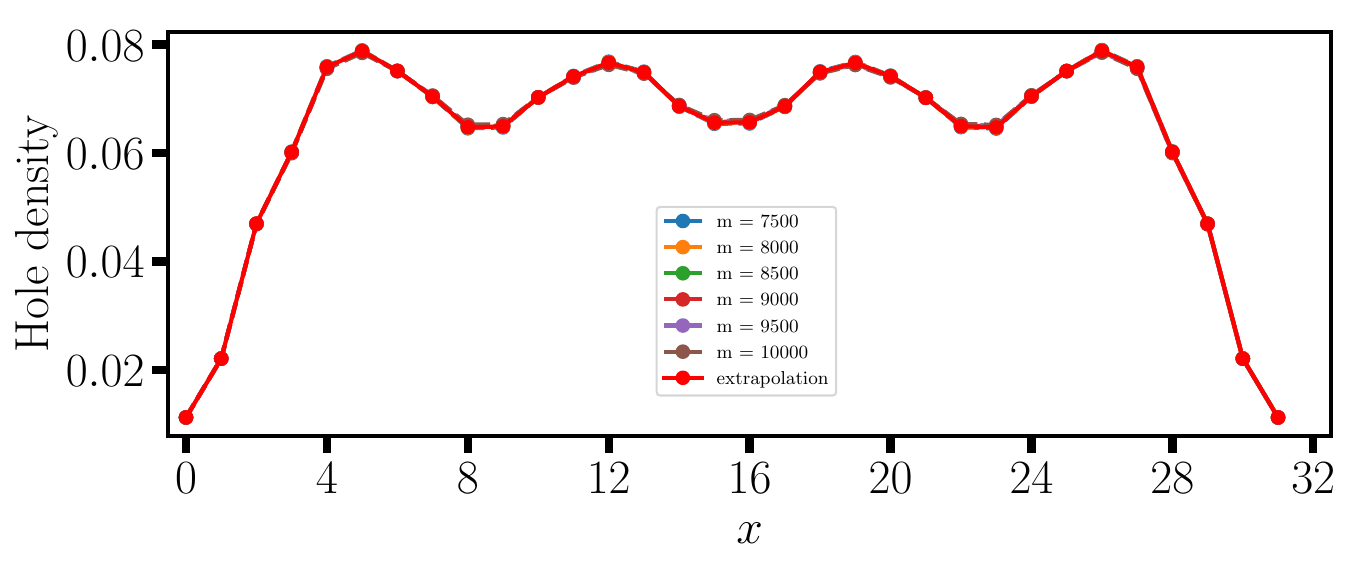}
	\caption{The staggered spin and hole density for the $4 \times 32$, $U = 8$, and $1/16$ doping system. Both DMRG results with finite kept states and result from an extrapolation
		with truncation error are shown. The dashed horizontal line in the upper panel represents $0$.} 
	\label{spin_hole_scale_4_32}
\end{figure}

\begin{figure}[b]
	\includegraphics[width=80mm]{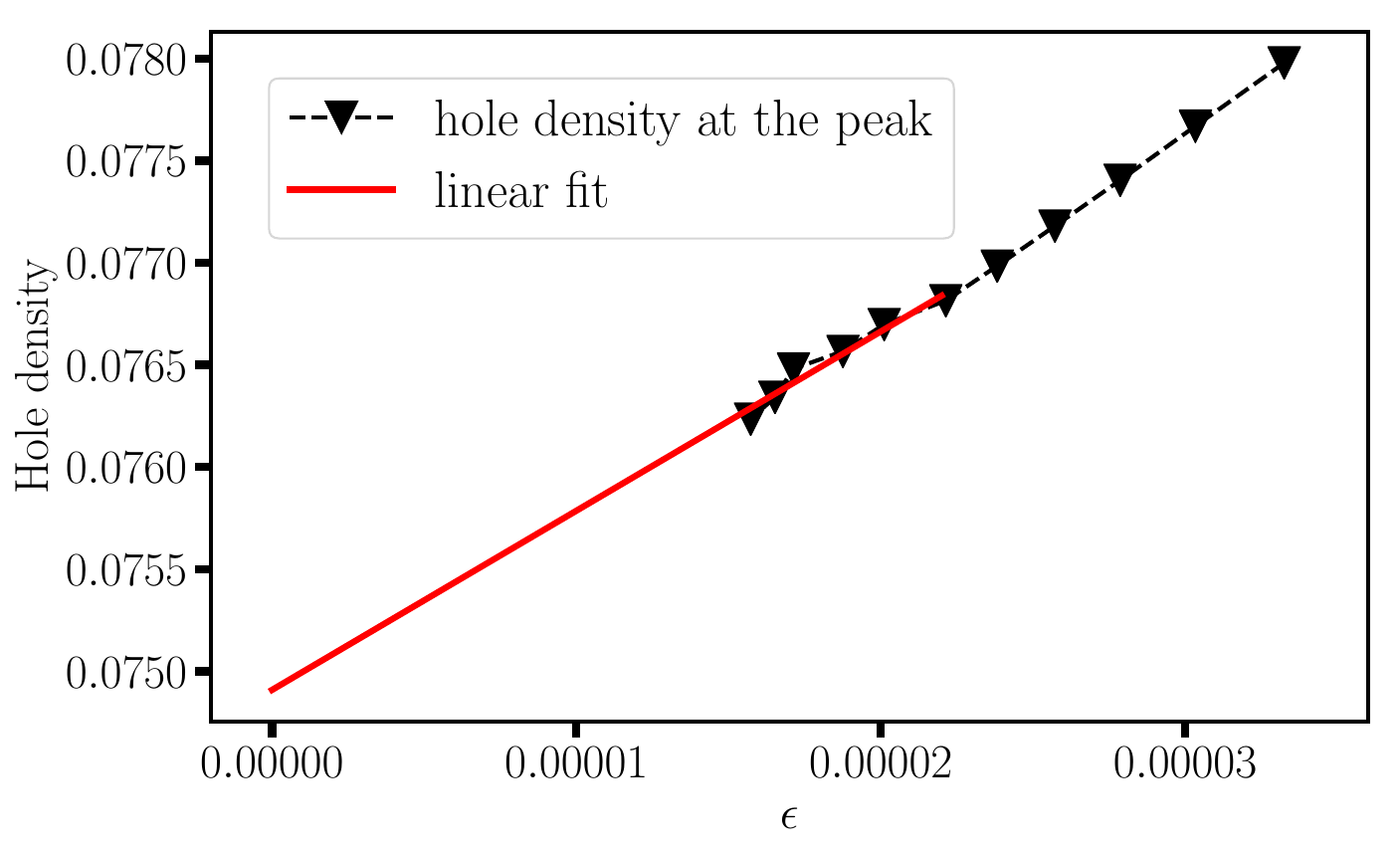}
	\caption{Scaling of the peak value of hole density in Fig.~\ref{spin_hole_scale_4_32} with truncation error in DMRG calculation.} 
	\label{peak_scale}
\end{figure}

\begin{figure}[t]
	\includegraphics[width=80mm]{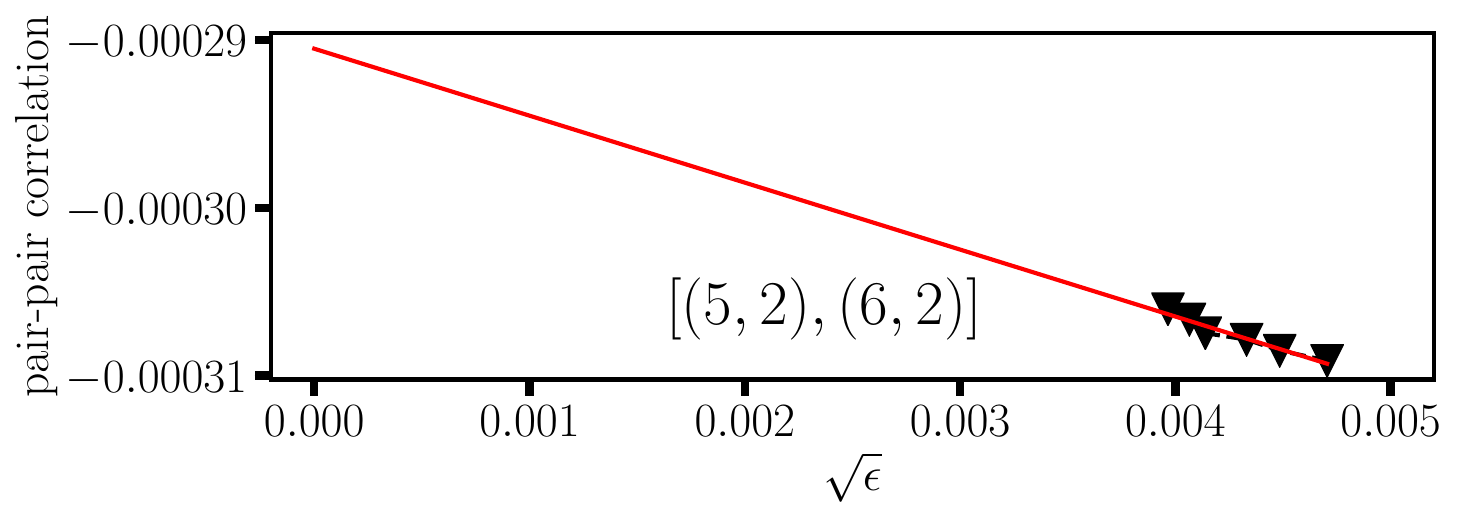}
	\includegraphics[width=80mm]{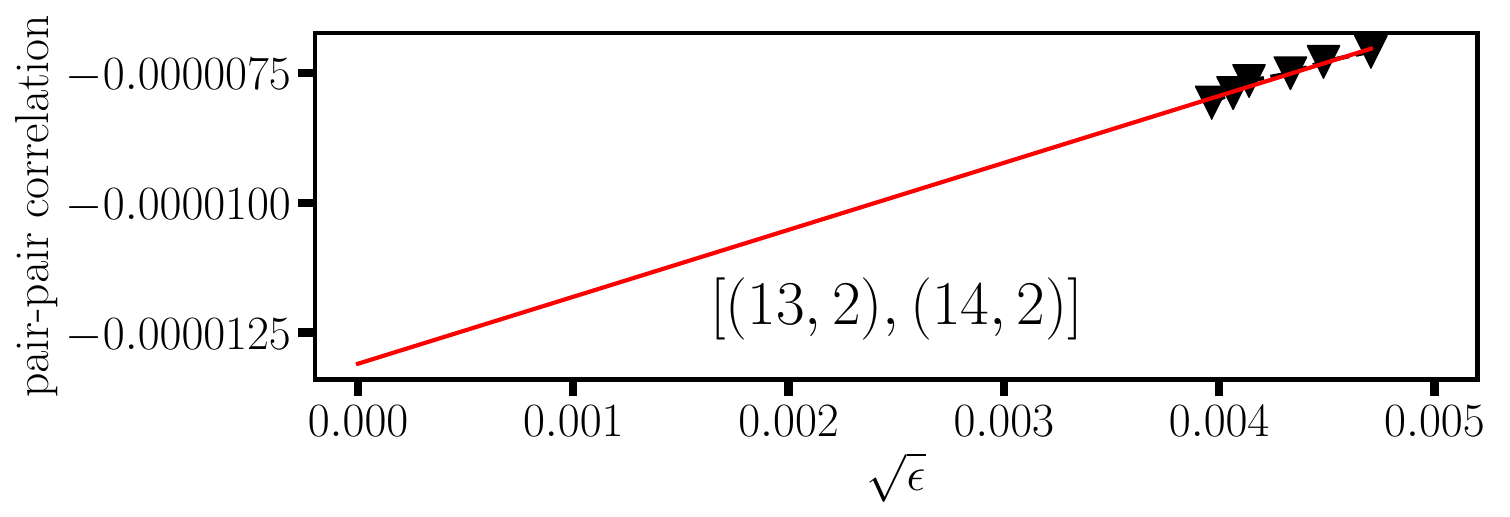}
	\includegraphics[width=80mm]{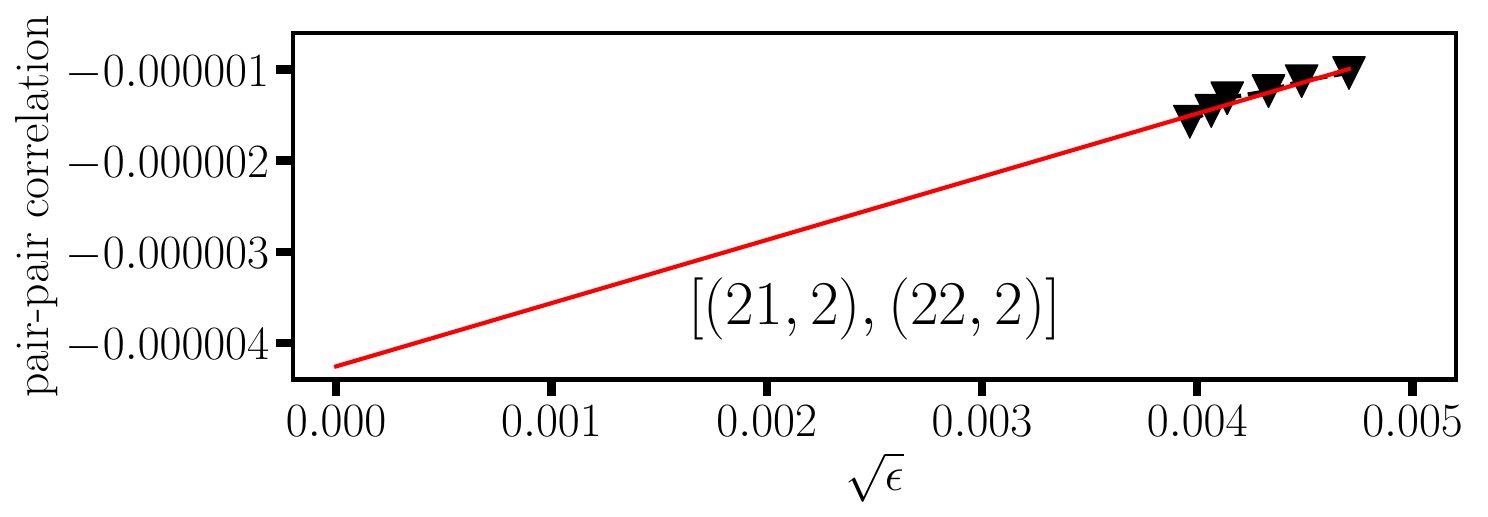}
	\caption{Scaling of the pair-pair correlation for bonds [(5,2), (6,2))], [(13,2), (14,2))], and [(21,2), (22,2))] with the square root of truncation error
		for the $4 \times 32$, $U = 8$, and $1/16$ doping system in DMRG calculation. The reference bond is placed at the
		edge between site $(1, 3)$ and $(2, 2)$.} 
	\label{pair_scale}
\end{figure}

In Fig.~\ref{spin_hole_scale_4_32}, we show the spin and hole density for the $4\times 32$, system with different number of kept states in DMRG calculation.
in Fig.~\ref{peak_scale}, we show the extrapolation procedure for the peak hole density in Fig.~\ref{spin_hole_scale_4_32}.
In Fig.~\ref{pair_scale}, we show the extrapolation of pair-pair correlation for bonds [(5,2), (6,2))], [(13,2), (14,2))], and [(21,2), (22,2))] with the square root
of truncation error.

\begin{figure}[b]
	\includegraphics[width=80mm]{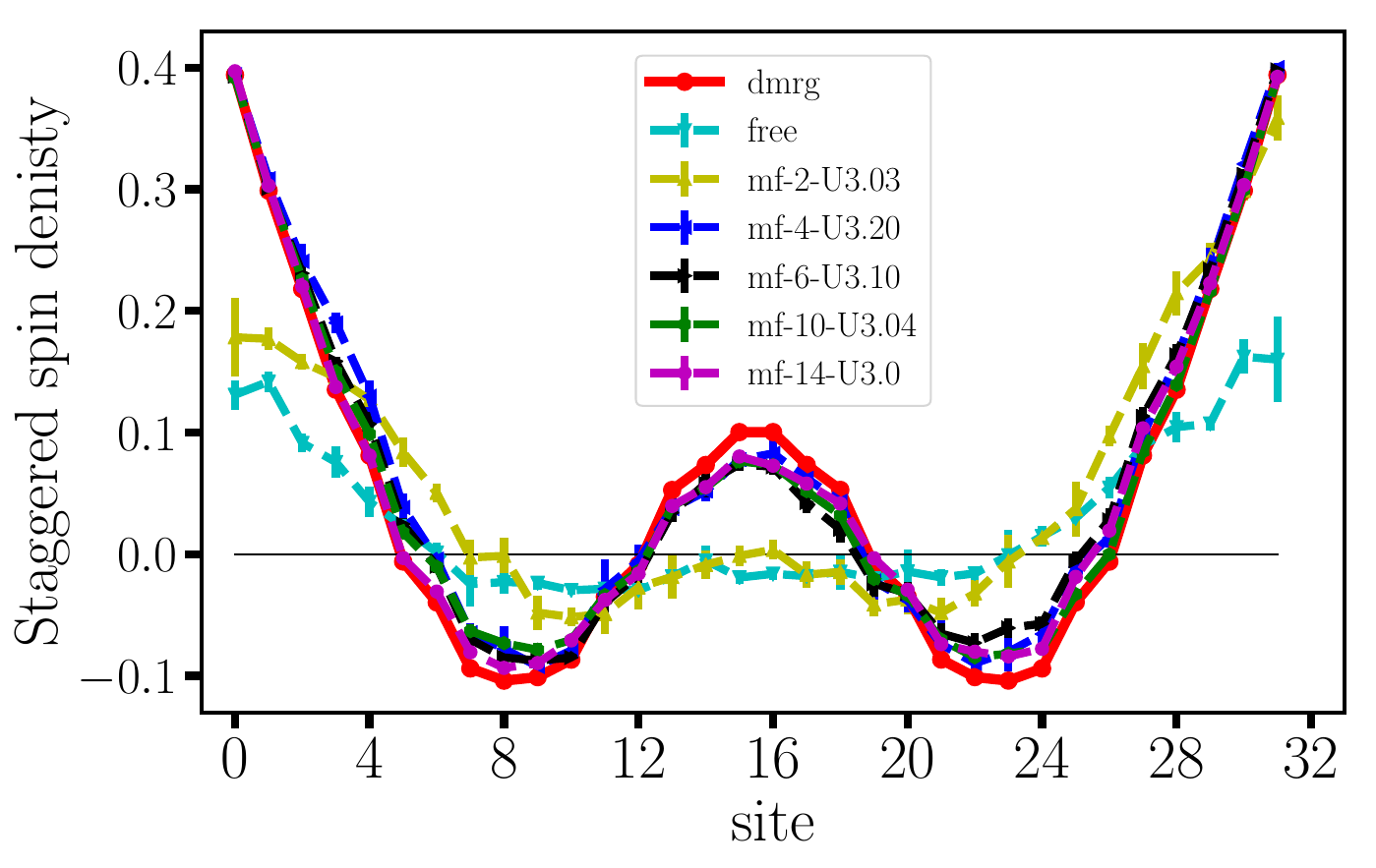}
	\includegraphics[width=80mm]{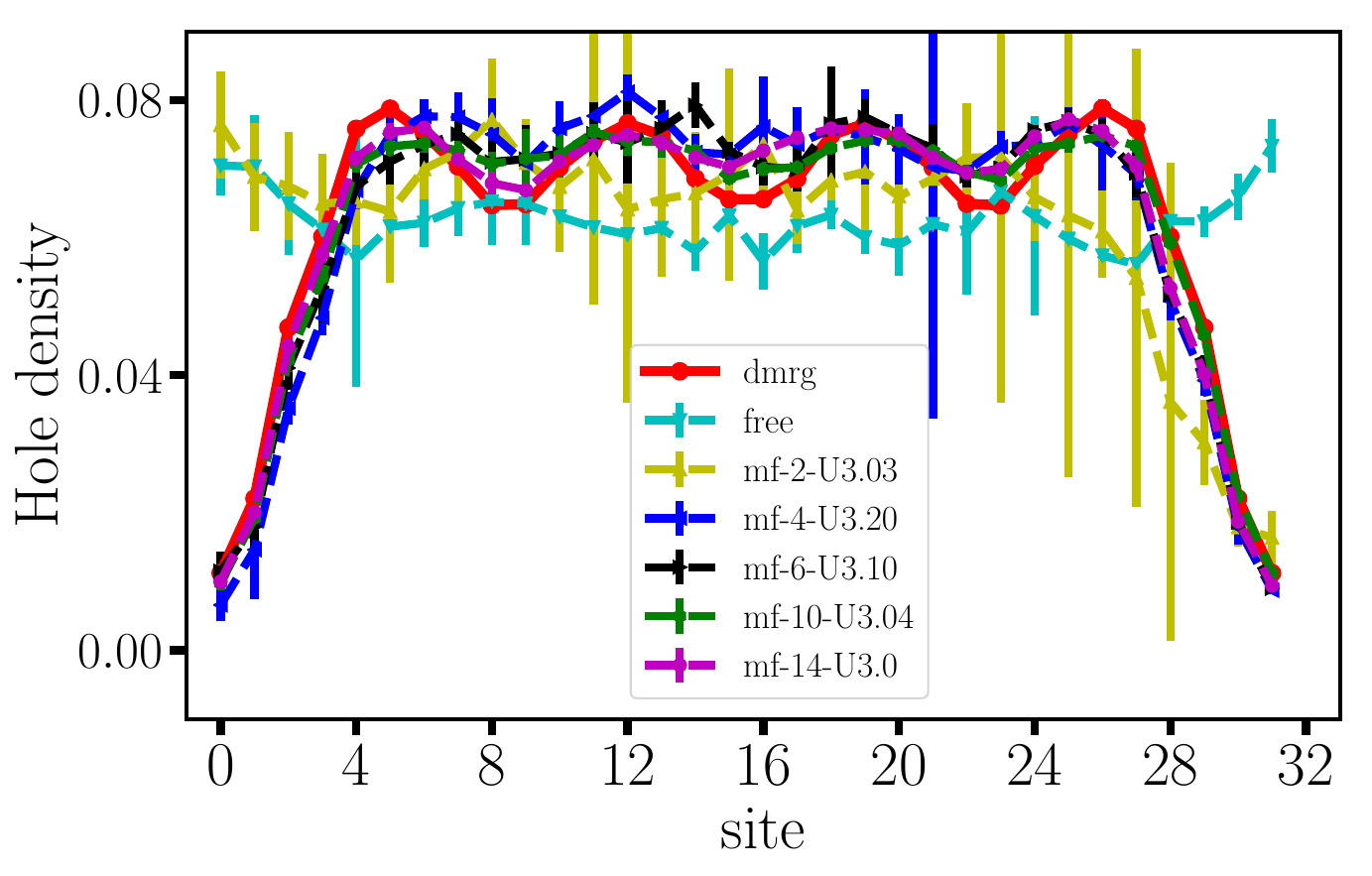}
	\caption{Self-consistent process in the CP-AFQMC calculation. The $U$ values in the label for each line are the effective $U$ values in
		the coupled mean-field Hamiltonian. The horizontal line in the upper panel represents $0$. We start the CP-AFQMC calculation with a free electron trial wave-function. The results after convergence
		agree well with the accurate DMRG results.} 
	\label{qmc_self}
\end{figure} 

\subsection{AFQMC with self-consistent optimized trial wave-function}
We couple the CP-AFQMC and a mean-field Hamiltonian to optimize the trial wave-function self-consistently as in \cite{PhysRevB.94.235119}.
in Fig.~\ref{qmc_self}, we show the convergence of spin and hole density in the self-consistent process for the $4\times 32$ system with $U = 8$ and 
$1/12$ doping. We start the CP-AFQMC calculation with a free electron trail wave-function, with which CP-AFQMC gives spin and hole density
far away from the accurate DMRG results. But the CP-AFQMC results gradually converges to the DMRG values within the self-consistent process. 
We notice that the effective interaction in the coupled mean-field Hamiltonian is $U_{eff} \approx 3$ after convergence.

\section{$U = 8$, $1/12$ doping results}
\begin{figure}[t]
	\includegraphics[width=80mm]{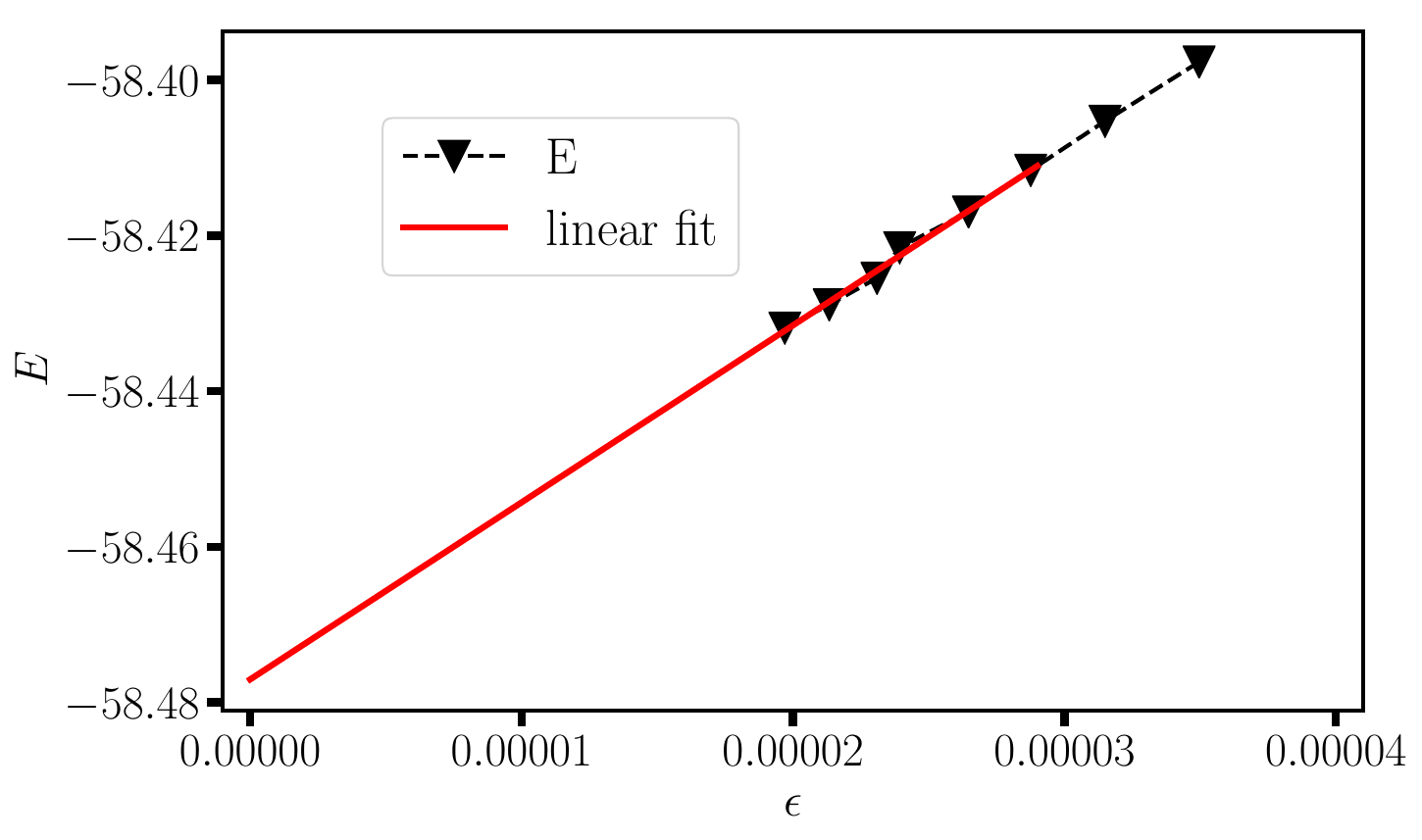}
	\caption{Scaling of energy with truncation error in DMRG calculation. The system is with size $4\times 24$, $U = 8$, and $1/12$ doping.
		The ground energy is estimated to be $-58.477(3)$ with a linear extrapolation using the $6$ data with smallest truncation errors.} 
	\label{E_scale_4_24}
\end{figure}

\begin{figure}[t]
	\includegraphics[width=80mm]{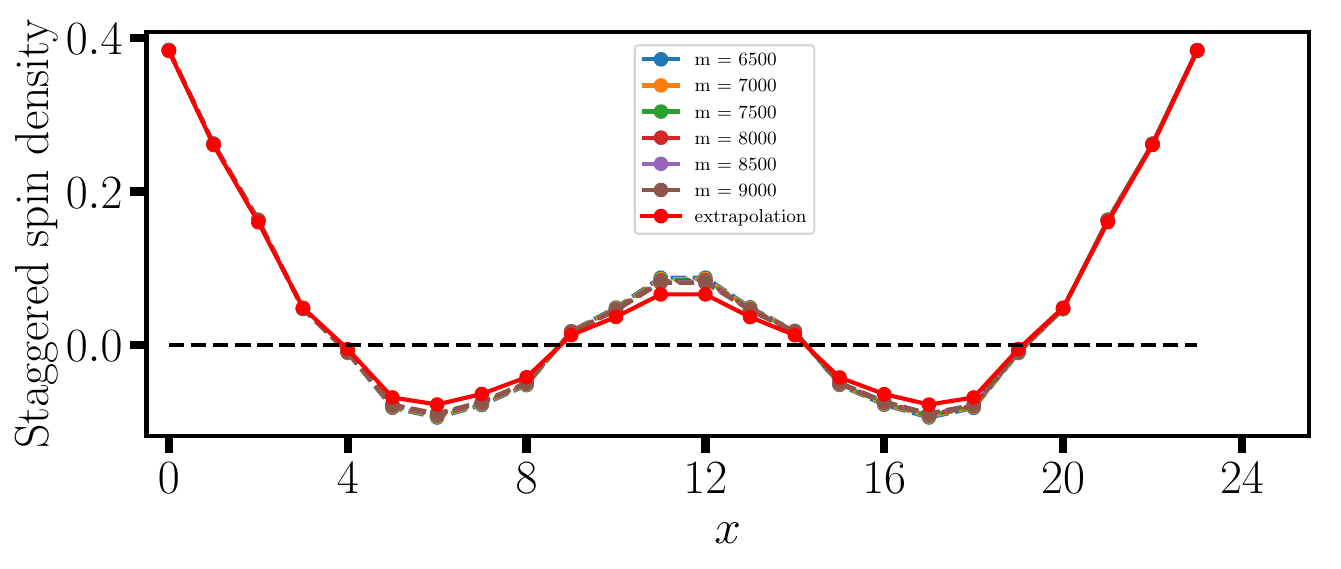}
	\includegraphics[width=80mm]{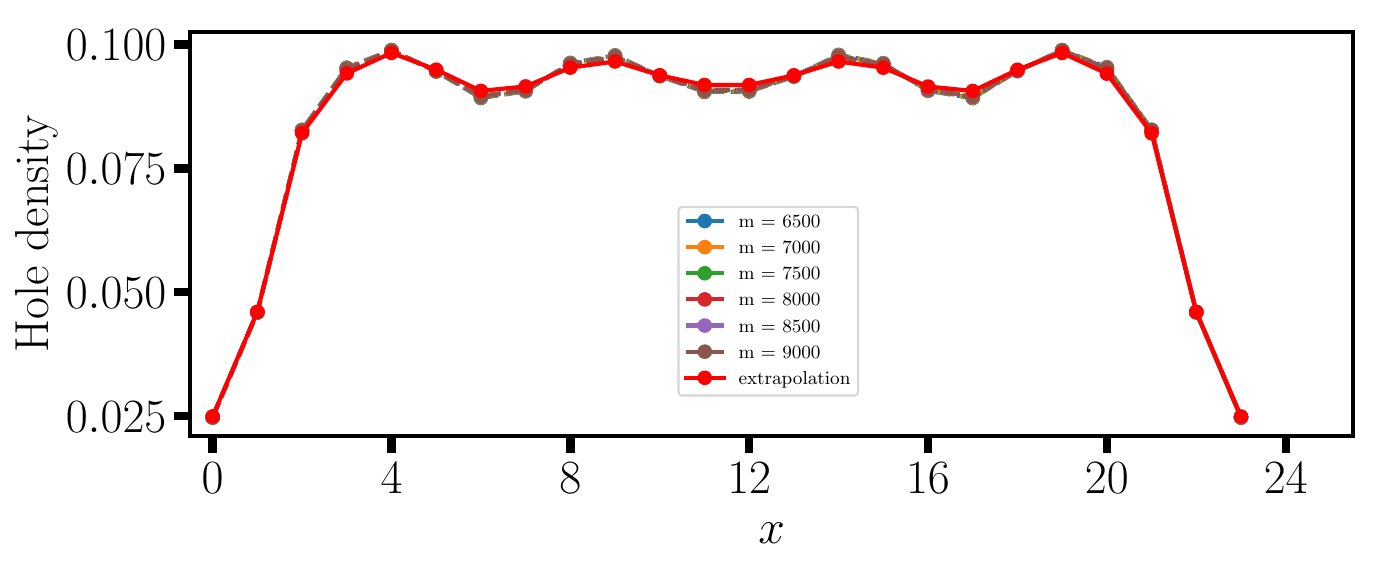}
	\caption{The staggered spin and hole density for the $4 \times 24$, $U = 8$, and $1/12$ doping system. Both DMRG results with finite kept states and result from an extrapolation
		with truncation error are shown. The dashed horizontal line in the upper panel represents $0$.} 
	\label{spin_hole_scale_4_24}
\end{figure}

\begin{figure*}[t]
	\includegraphics[width=59mm]{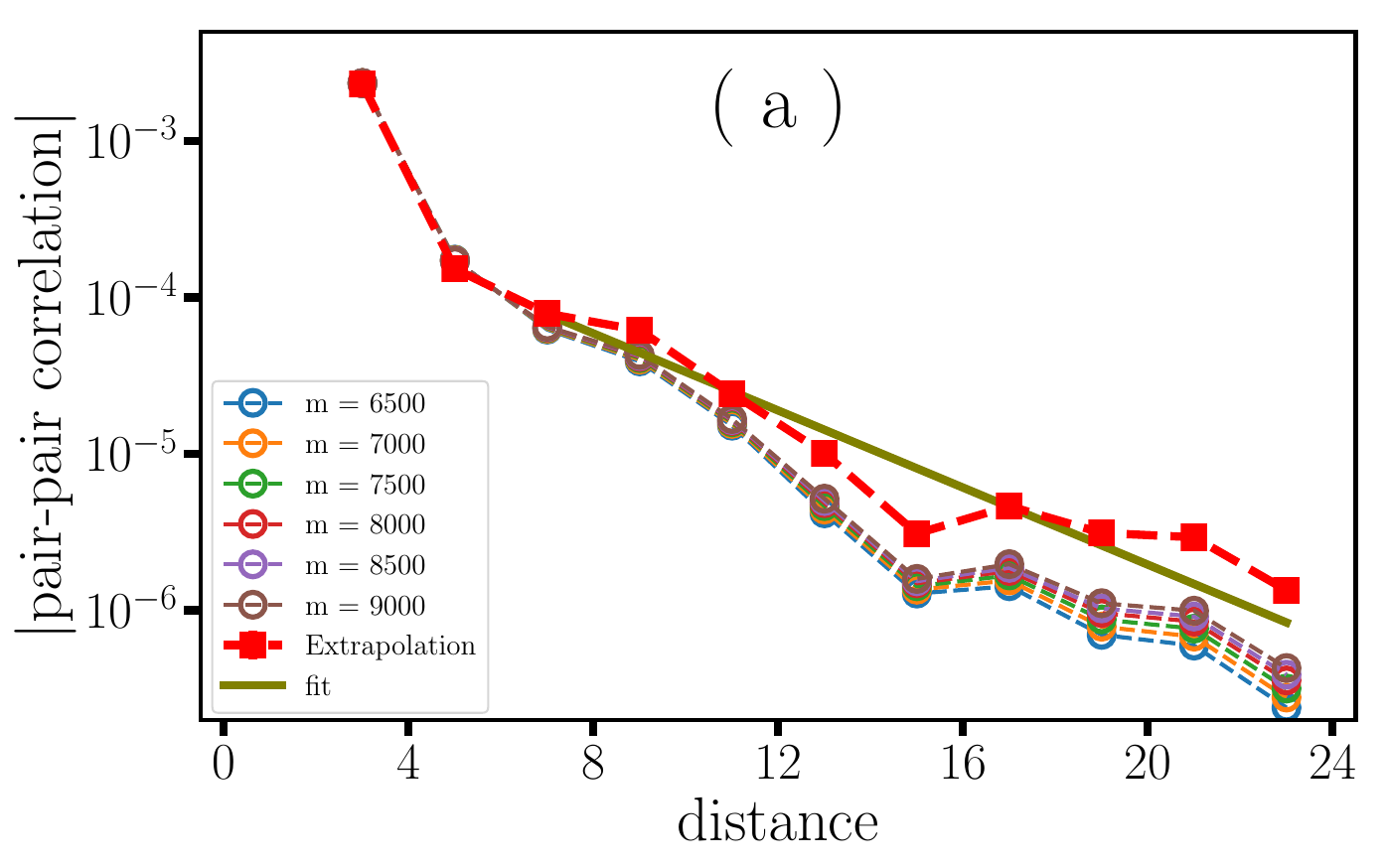}
	\includegraphics[width=59mm]{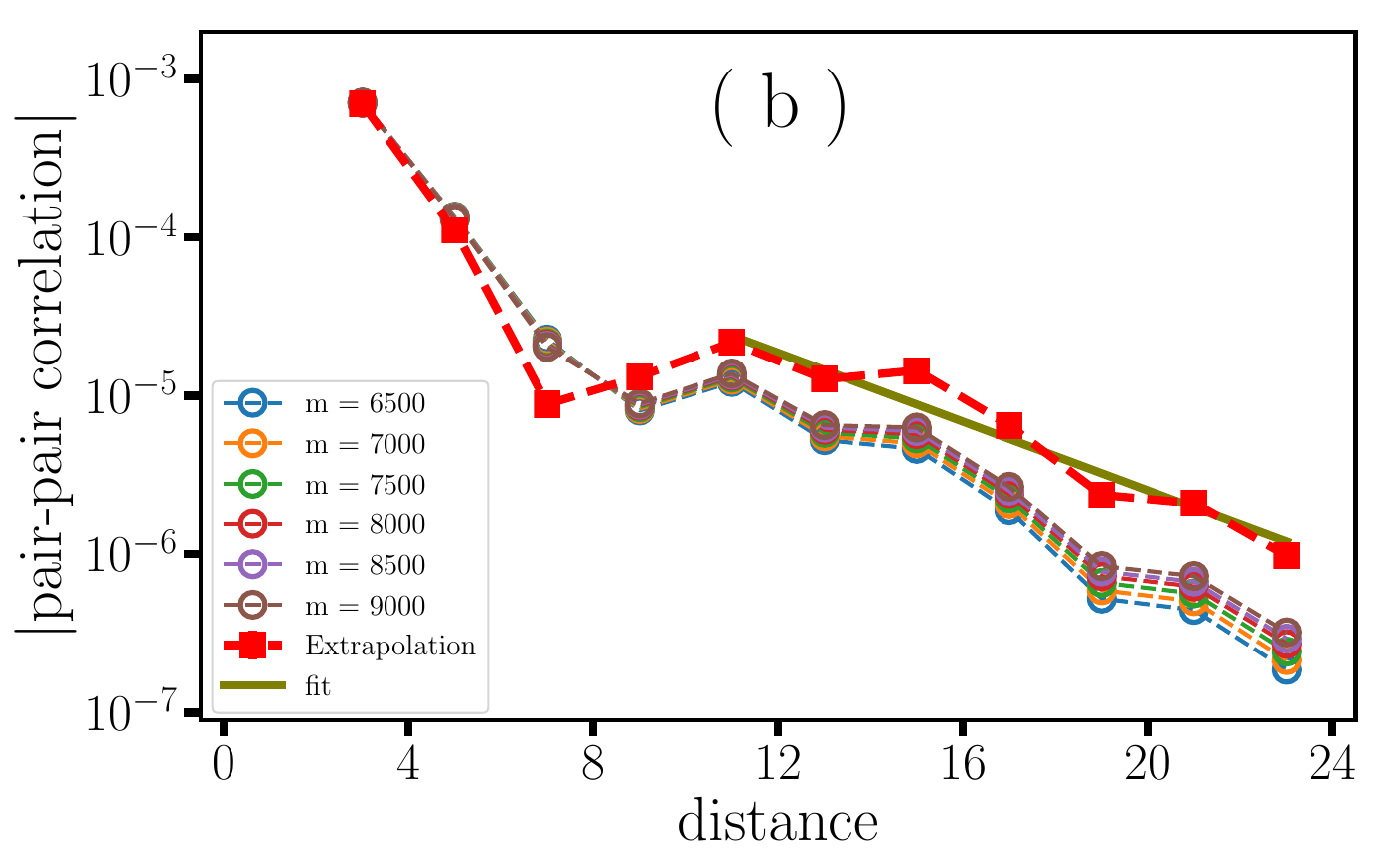}
	\includegraphics[width=59mm]{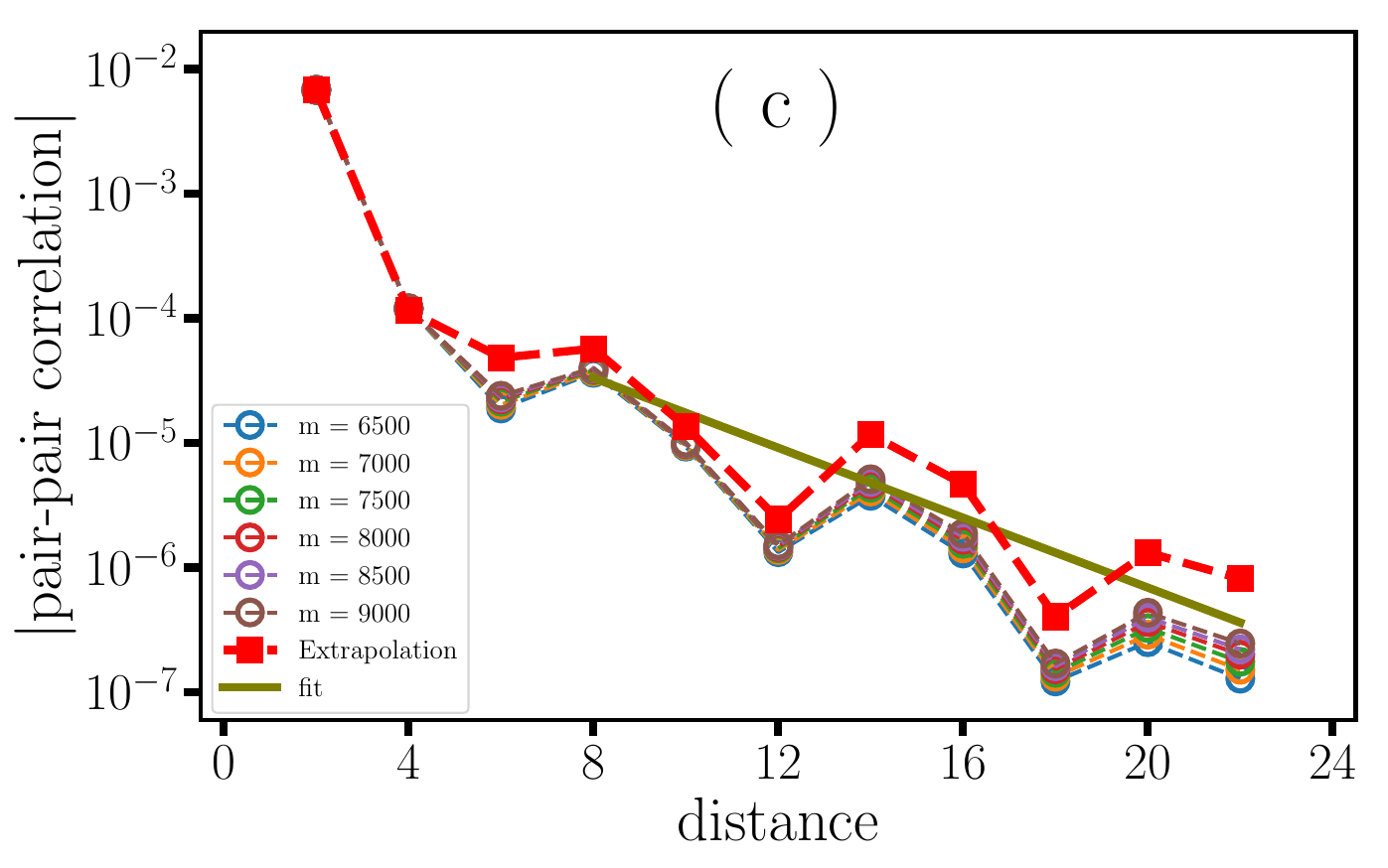}
	\caption{DMRG results for the absolute value of pair-pair correlations for the $4 \times 24$ system with $U = 8$ and $1/12$ doping. The reference bond is placed at the
		edge between site $(1, 3)$ and $(2, 2)$. Panels (a), (b), and (c) show the correlation versus distance between the reference bond (The A bond in the dashed oval
		in Fig.~\ref{lattice}) and the
		black (A), blue (B), and red (C) bonds (see Fig.~\ref{lattice}) respectively. Both results with finite kept states and from an extrapolation (red) with truncation errors are shown. An exponential decay can be seen from the fit (brown).}
	\label{pair_4_24}
\end{figure*}

\begin{figure*}[t]
	\includegraphics[width=180mm]{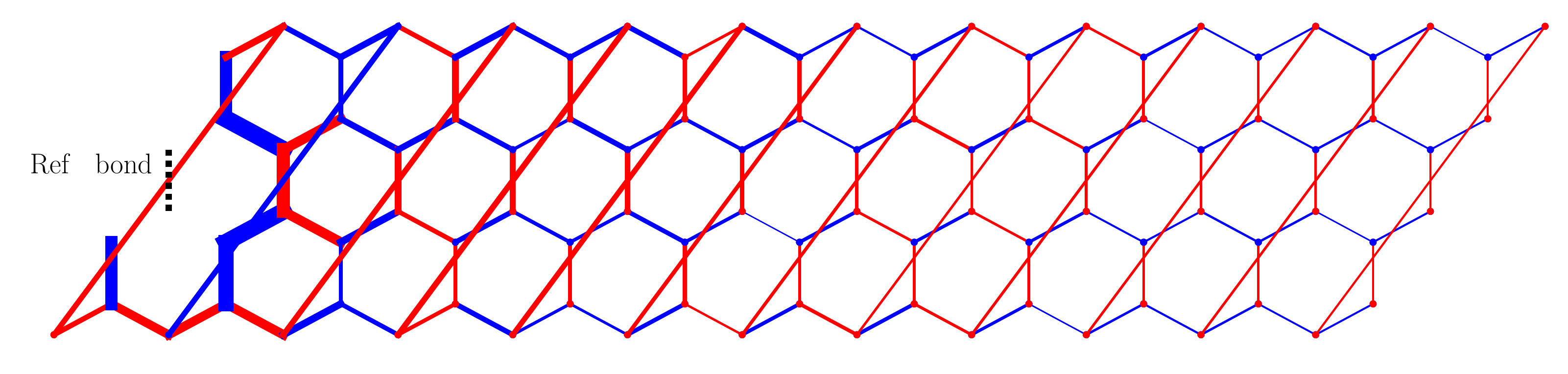}
	\caption{The pair-pair correlation pattern on the whole lattice for the $4 \times 24$ system with $U = 8$ and $1/12$ doping . The reference bond is denoted by the dashed
		black line. Red (blue) color
		means positive (negative) correlation values. The thickness of each bond is proportional to  $\langle \hat{\Delta}_{i'j'}^{\dagger} \hat{\Delta}_{ij} \rangle ^\frac{1}{4}$ 
		to make the line visible.
		We can see that the A (B) bonds (definition in Fig.~\ref{lattice}) have positive (negative) correlation at large distance, while the sign of C bonds oscillate with
		the distance to the reference bond. Similar as Fig.~\ref{pair_color} in the main text.} 
	\label{pair_4_24_sign}
\end{figure*}

\begin{figure}[t]
	\includegraphics[width=80mm]{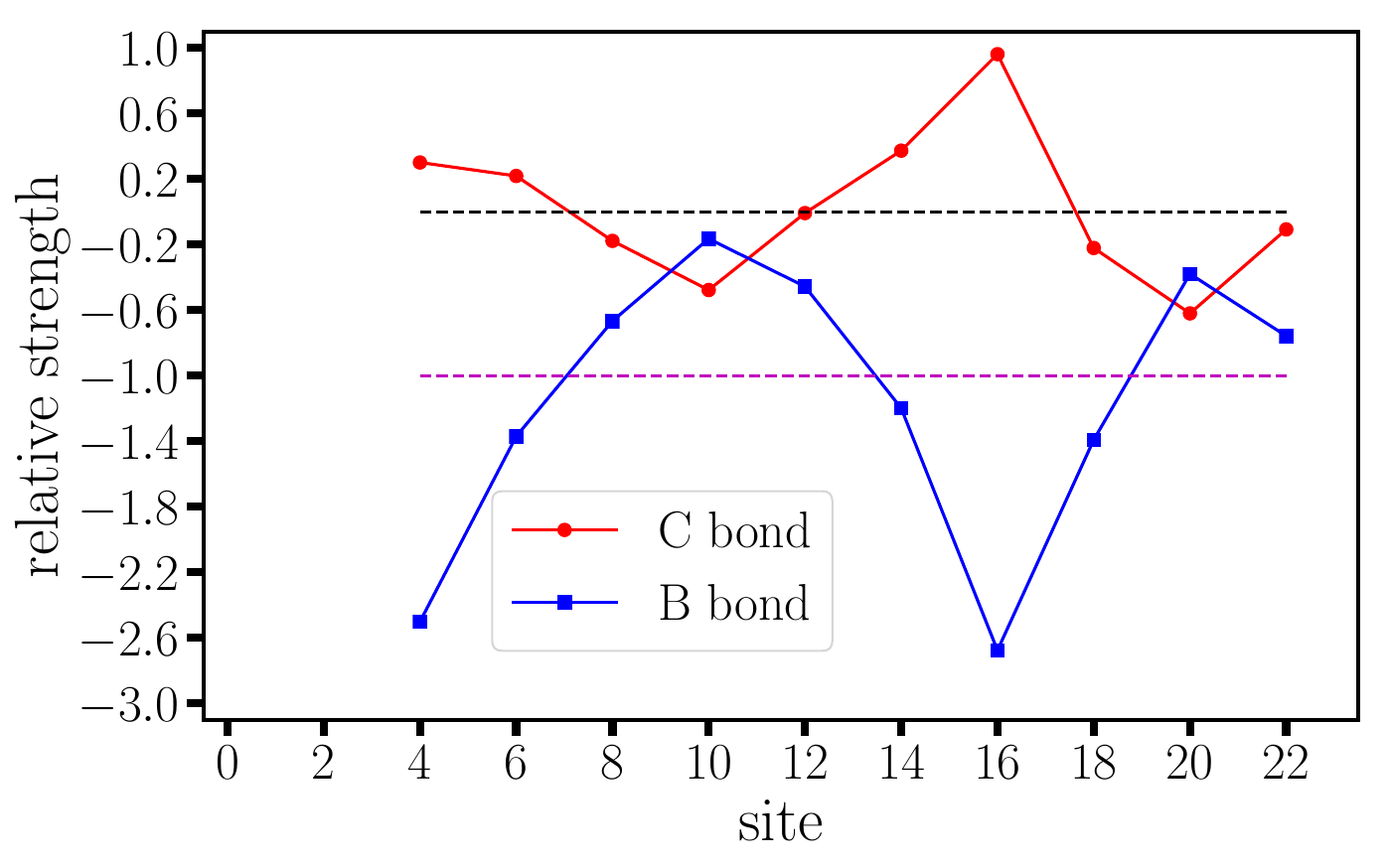}
	\caption{Relative strength of the pair-pair correlation function for or the $4 \times 24$, $U = 8$, and $1/12$ doping system.
		The dashed horizontal lines represent $0$ and $-1$. Similar as Fig.~\ref{pair_ratio} in the main text.} 
	\label{pair_4_24_ratio}
\end{figure}

In this section, we show the results for the  $4 \times 24$ system with $U = 8$ and $1/12$ doping.
AF magnetic pining fields with strength $|h_m| = 0.5$ are applied at the open edges.
In Fig.~\ref{E_scale_4_24},  we show the scaling of energy with the truncation error in DMRG calculation. In Fig.~\ref{spin_hole_scale_4_24},
we show the spin and hole density for this system. We can find a half-filled stripe state for the ground state. We plot the 
pair-pair correlation function in Fig.~\ref{pair_4_24}, from which we can see an exponential decay. In Fig.\ref{pair_4_24_sign}
and Fig.\ref{pair_4_24_ratio} we show the pair-pair correlation pattern for the same system. 
The sign structure is similar as the $1/16$ doping case, i.e., the A (B) bonds are all positive (negative), while the sign of C bonds
oscillates. The relative strength in Fig.\ref{pair_4_24_ratio} shows the local $d$-wave structure is
not as clear as the $4 \times 32$, $1/16$ doping system.  

\section{$U = 8$, $1/8$ doping results}
\begin{figure}[t]
	\includegraphics[width=70mm]{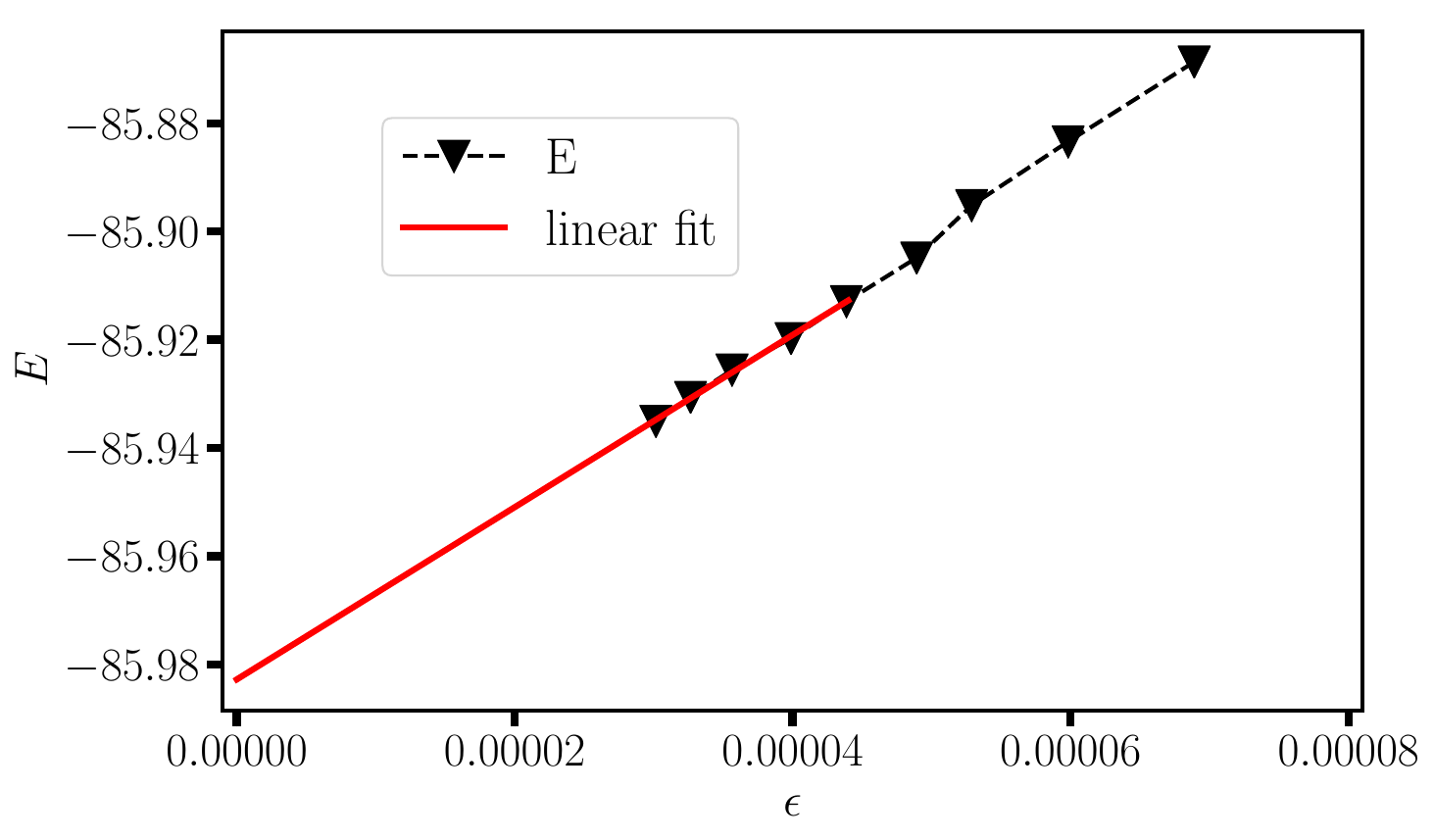}
	\caption{Scaling of energy with truncation error in DMRG calculation. The system is with size $4\times 32$, $U = 8$ and $1/8$ doping.
		The ground state energy is estimated to be $-85.986(3)$ with a linear extrapolation using the $6$ data with smallest truncation errors.} 
	\label{E_scale_4_32_1-8}
\end{figure}

\begin{figure}[t]
	\includegraphics[width=80mm]{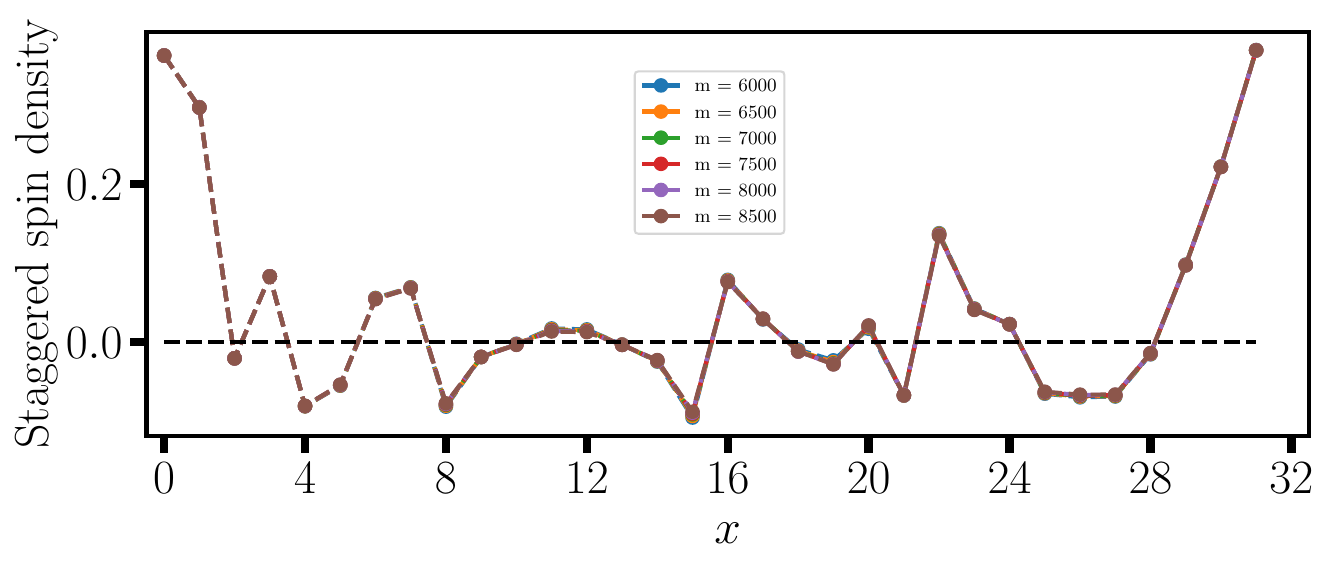}
	\includegraphics[width=80mm]{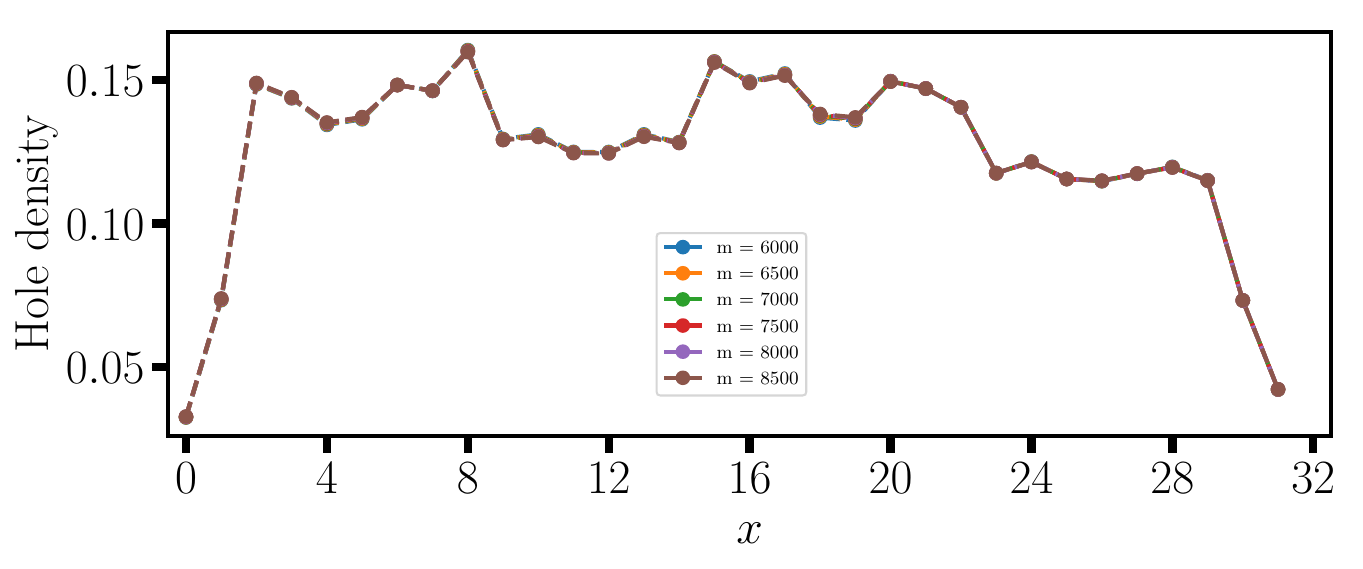}
	\caption{The staggered spin and hole density for the $4 \times 32$, $U = 8$, and $1/8$ doping system. The dashed horizontal line in the upper panel represents $0$. } 
	\label{spin_hole_scale_4_32-1-8}
\end{figure}

\begin{figure*}[t]
	\includegraphics[width=59mm]{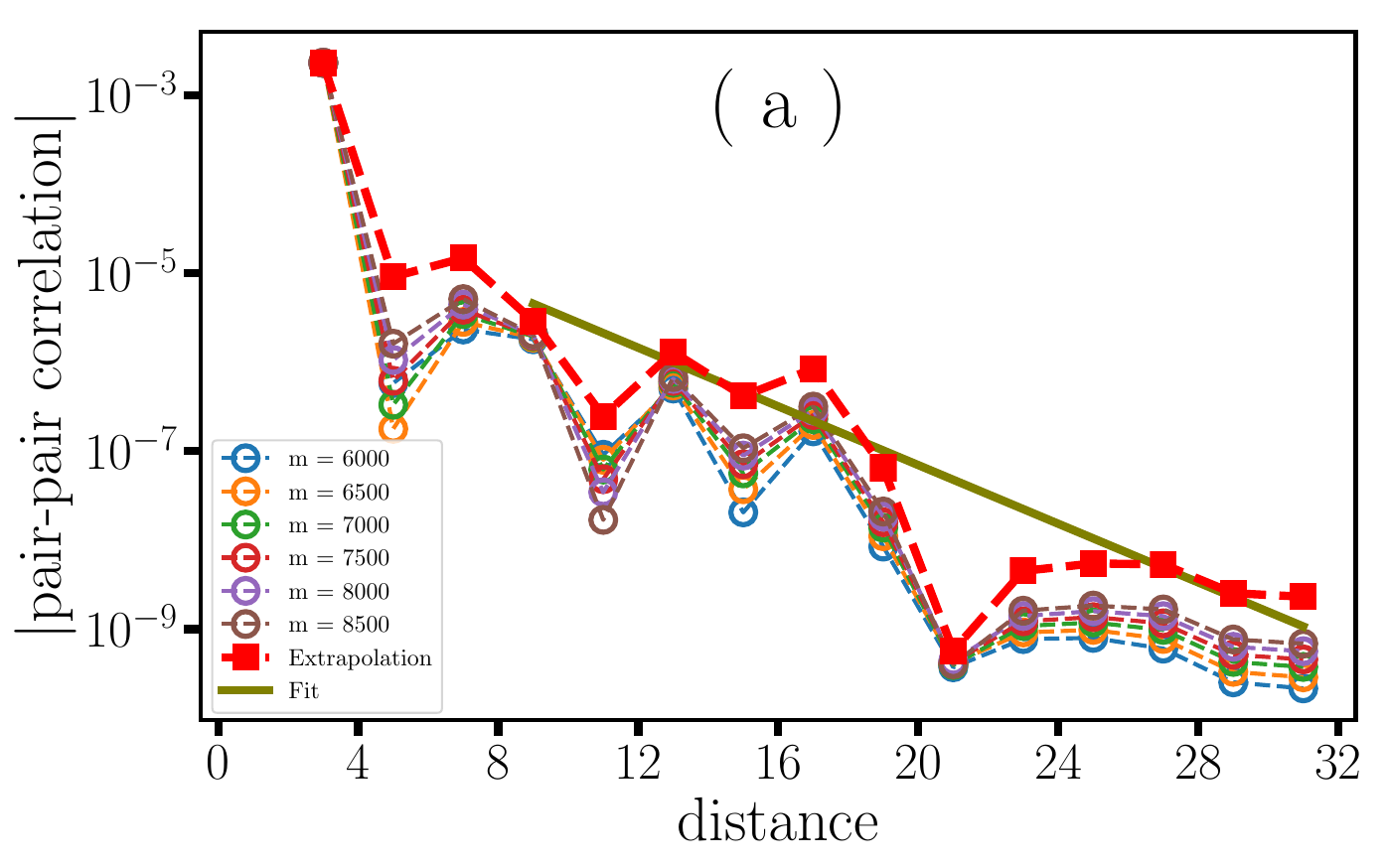}
	\includegraphics[width=59mm]{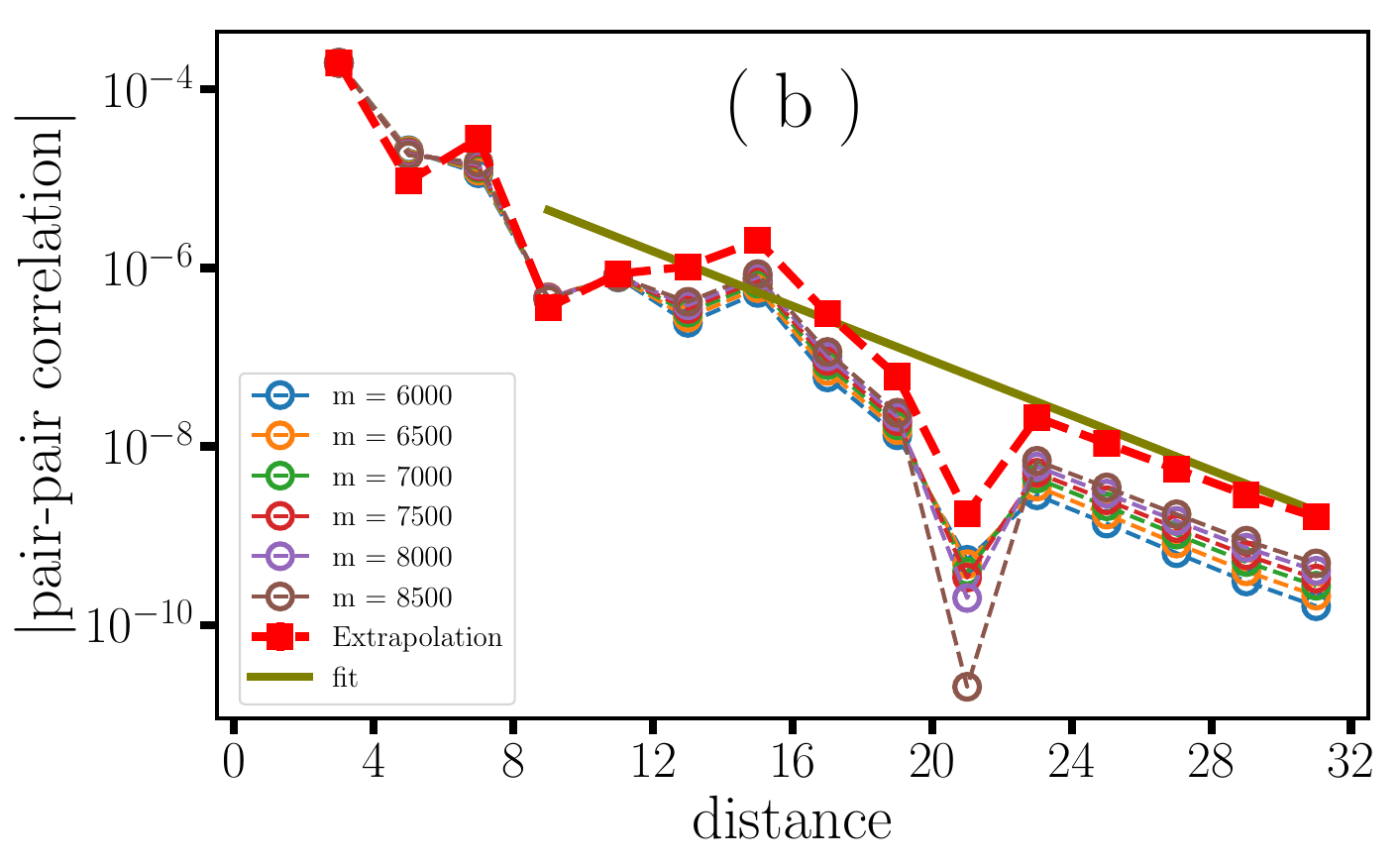}
	\includegraphics[width=59mm]{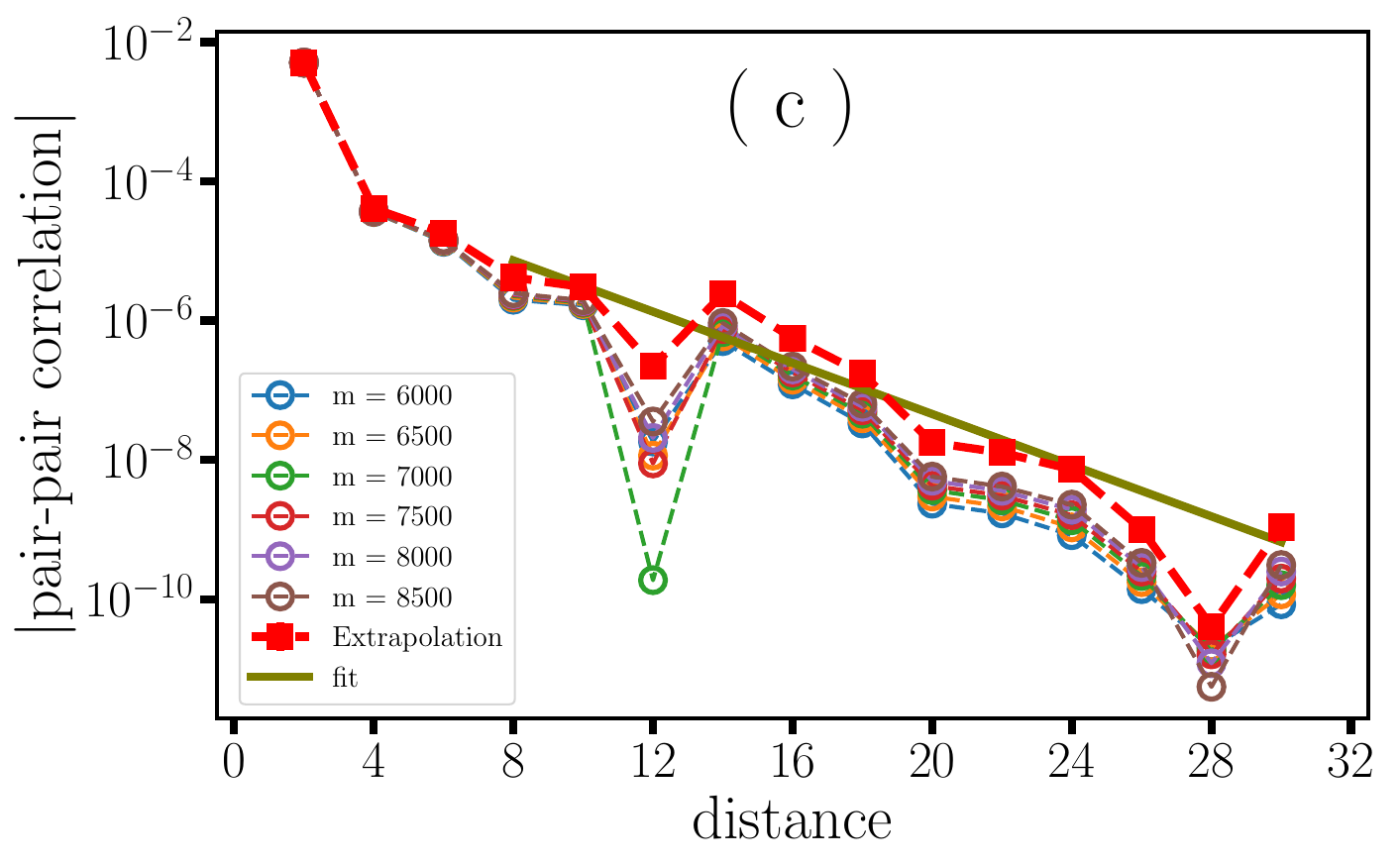}
	\caption{DMRG results for the absolute value of pair-pair correlations for $4 \times 32$ system with $U = 8$ and $1/8$ doping. The reference bond is placed at the
		edge between site $(1, 3)$ and $(2, 2)$. Panels (a), (b), and (c) show the correlation versus distance between the reference bond (The A bond in the dashed oval
		in Fig.~\ref{lattice}) and the
		black (A), blue (B), and red (C) bonds (see Fig.~\ref{lattice}) respectively. Both results with finite kept states and from an extrapolation (red) with truncation
		errors are shown. An exponential decay can be seen from the fit (brown).}
	\label{pair_4_32-1-8}
\end{figure*}

\begin{figure*}[t]
	\includegraphics[width=180mm]{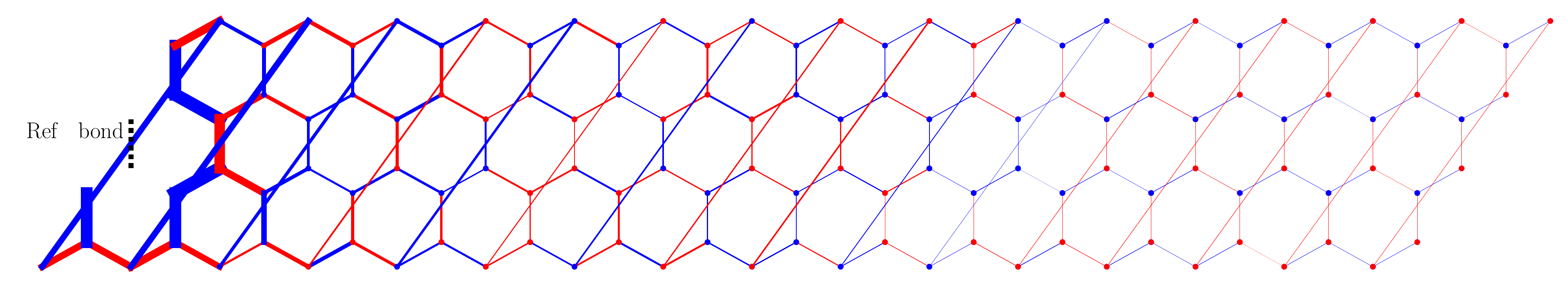}
	\caption{The pair-pair correlation pattern on the whole lattice for the $4 \times 32$ system with $U = 8$ and $1/8$ doping . The reference bond is denoted by the dashed
		black line. Red (blue) color
		means positive (negative) correlation values. The thickness of each bond is proportional to  $\langle \hat{\Delta}_{i'j'}^{\dagger} \hat{\Delta}_{ij} \rangle ^\frac{1}{4}$ 
		to make the line visible.
		We can see that the sign of A, B, and C bonds all oscillate with
		the distance to the reference bond. Similar as Fig.~\ref{pair_color} in the main text.} 
	\label{pair_4_32_sign-1-8}
\end{figure*}

In this section, we show the results for the $4 \times 32$ system with $U = 8$ and $1/8$ doping.
AF magnetic pining fields with strength $|h_m| = 0.5$ are applied at the open edges.
In Fig.~\ref{E_scale_4_32_1-8},  we show the scaling of energy with the truncation error in DMRG calculation. In Fig.~\ref{spin_hole_scale_4_32-1-8},
we show the spin and hole density for this system. We find no stripe state at $1/8$ doping. We plot the 
pair-pair correlation function in Fig.~\ref{pair_4_32-1-8}, from which we can clearly see an exponential decay.
We also notice the pair-pair correlation is weaker than the $1/8$ and $1/12$ doping cases. In Fig.~\ref{pair_4_32_sign-1-8}, we show the pair-pair
correlation pattern for the same system. We find that the sign of A, B, and C bonds all oscillate with the distance which indicate no local d-wave pattern
at $1/8$ doping.  

\end{document}